\documentclass[aps,pre,reprint,superscriptaddress, amsmath, amssymb, showpacs]{revtex4-1}

\usepackage{appendix}

 \usepackage{gensymb}
 \usepackage{subcaption}
 \usepackage{caption}
\newcommand*{\rom}[1]{\expandafter\@slowromancap\romannumeral #1@}
\newcommand{\vc}{\mathbf}
\usepackage{graphicx}
\usepackage{color}
\usepackage{xcolor}
\usepackage{comment} 

\begin{document}

\title{Temperature Relaxation Rates in Strongly Magnetized Plasmas}


\author{Louis Jose}
\affiliation{Nuclear Engineering \& Radiological Sciences, University of Michigan, Ann Arbor, Michigan 48109, USA}

\author{James C. Welch III}
\affiliation{Nuclear Engineering \& Radiological Sciences, University of Michigan, Ann Arbor, Michigan 48109, USA}

\author{Timothy D. Tharp}
\affiliation{Physics, Marquette University, Milwaukee, Wisconsin, 53233, USA}

\author{Scott D. Baalrud}
\email{baalrud@umich.edu} 
\affiliation{Nuclear Engineering \& Radiological Sciences, University of Michigan, Ann Arbor, Michigan 48109, USA}


\date{\today}

\begin{abstract}
Strongly magnetized plasmas, characterized by having a gyrofrequency larger than the plasma frequency ($\beta = \omega_c/\omega_p \gg 1$), are known to exhibit novel transport properties. Previous works studying pure electron plasmas have shown that strong magnetization significantly inhibits energy exchange between parallel and perpendicular directions, leading to a prolonged time for relaxation of a temperature anisotropy. Recent work studying repulsive electron-ion interactions showed that strong magnetization increases both the parallel and perpendicular temperature relaxation rates of ions, but in differing magnitudes, resulting in the formation of temperature anisotropy during equilibration. This previous study treated electrons as a heat bath and assumed weak magnetization of ions. Here, we broaden this analysis and compute the full temperature and temperature anisotropy evolution over a broad range of magnetic field strengths. 
It is found that when electrons are strongly magnetized ($\beta_e \gg 1$) and ions are weakly magnetized ($\beta_i \ll 1$), the magnetic field strongly suppresses the perpendicular energy exchange rate of electrons, whereas the parallel exchange rate slightly increases in magnitude compared to the value at weak magnetization. In contrast, the ion perpendicular and parallel energy exchange rates both increase in magnitude compared to the values at weak magnetization. Consequently,  equilibration causes the electron parallel temperature to rapidly align with the ion temperature, while the electron perpendicular temperature changes much more slowly. 
It is also shown that when both ions and electrons are strongly magnetized ($\beta_i, \beta_e \gg 1$) the ion-electron perpendicular relaxation rate dramatically decreases with magnetization strength as well. 
\end{abstract}




\maketitle





\section{Introduction}

Plasmas are frequently in a non-equilibrium state with electrons and ions at different temperatures. 
In magnetized plasmas, it is also common to have different temperatures parallel and perpendicular to the magnetic field~\cite{Ott_PRE_2017}. Here, we study how plasma equilibrates from this non-equilibrium state through Coulomb collisions when it is strongly magnetized. Strongly magnetized plasmas are defined by having an electron gyrofrequency that exceeds the electron plasma frequency~\cite{Baalrud_PRE_2017,ott2011diffusion, Kahlert_Phys_Rev_Res_2022, Kahlert_CPP_2023}, i.e., $\beta_e  = \omega_{ce}/\omega_{pe} >~1$, where $\omega_{ce} = eB/m_ec$ and $\omega_{pe} = \sqrt{4 \pi e^2 n_e/m_e}$. Understanding how temperature relaxation is modified by strong magnetization has applications in many experiments, such as antimatter traps~\cite{Fajans_POP_2020,Stenson_JPP_2017,Ahmadi_nature_comm_2017,Baker_Nature_Comm_2021}, magnetized dusty plasmas~\cite{Thomas_PPCF_2012}, non-neutral plasma~\cite{Beck_PRL_1992, Glinsky_Phys_Fluids_1992}, magnetized ultracold neutral plasmas~\cite{Gorman_PRA_2022,Gorman_PRL_2021,Zhang_PRL_2008,Sprenkle_PRE_2022,Guthrie_POP_2021,Pak_PRE_2024}, and pinch experiments~\cite{Bennett_Phys_Rev_Accel_Beams_2021}, in addition to being an interesting fundamental physics problem.
For example, previous works studying pure electron plasmas have shown that strong magnetization significantly inhibits energy exchange between parallel and perpendicular directions, leading to prolonged time for the relaxation of temperature anisotropy~\cite{Glinsky_Phys_Fluids_1992, Beck_PRL_1992}.

Here, a recently developed kinetic theory for strongly magnetized plasmas~\cite{Jose_POP_2020} is applied to model the evolution of temperature and temperature anisotropy of an electron-ion plasma interacting through a repulsive Coulomb potential. 
It is found that when the electrons are strongly magnetized ($\beta_e \gg 1)$ but ions are weakly magnetized ($\beta_i \ll 1$), the ion parallel and perpendicular energy exchange densities increase in magnitude compared to the weakly magnetized ($\beta_e < 1$) limit. However, the electron perpendicular energy exchange rate is strongly suppressed by the strong magnetization of electrons. 
As energy conservation requires, the parallel electron energy exchange rate increases correspondingly to match the total ion energy exchange rate. 
This leads to a three step process in the total temperature evolution. 
The fastest timescale is associated with ion-ion interactions, which prevents any significant ion temperature anisotropy from forming. 
The second timescale is associated with the relaxation of the parallel electron temperature with the ion temperature. 
The third and much longer timescale is associated with the relaxation of the perpendicular electron temperature with the others. 
Because it is so slow, the perpendicular electron temperature relaxation sets the overall temperature relaxation rate.

This analysis is relevant to modeling antiproton temperature evolution in the Antihydrogen Laser Physics Apparatus (ALPHA)~\cite{Ahmadi_nature_comm_2017,Baker_Nature_Comm_2021}. The ALPHA experiment synthesizes antihydrogen from antiprotons and positrons and makes precision measurements of antimatter to test fundamental symmetries. Before antihydrogen can be synthesized, nonneutral plasmas must be prepared.  In one stage of this preparation, antiprotons are cooled collisionally with electrons in a Penning trap. The magnetization of these plasmas corresponds to a $\beta_e$ value of a few hundred, a regime in which the electrons are strongly magnetized but the ions are not. 
This work focuses on modeling the antiproton-electron temperature relaxation, so only repulsive interactions are considered. It should be noted that our previous work~\cite{Jose_POP_2022, Jose_POP_2023} showed significant differences between the attractive and repulsive cases. So the results shown here are not expected to apply directly to attractive electron-ion interactions, which will be the subject of future work. 

Traditionally, the temperature relaxation rate is computed from plasma kinetic theories such as those based on the Boltzmann equation~\cite{ferziger1972mathematical}, Fokker Planck equation~\cite{Rosenbluth_PR_1957}, or Lennard-Balescu equation~\cite{Lenard_AP_1960, Balescu_Phys_Fluids_1960}. 
A result of these models that is that temperature relaxation is described by a single Coulomb collision rate. 
However, these are restricted to weakly magnetized plasmas because the expansion parameters they are based on are related to the gyroradius being larger than the Debye length. 
In this work, we use a recent generalized kinetic theory that treats strongly magnetized plasmas~\cite{Jose_POP_2020,Jose_POP_2021}. This generalizes the Boltzmann collision operator to account for the gyromotion of particles during Coulomb collisions. 
It also accounts for aspects of strong Coulomb coupling by modeling interactions using the potential of mean force~\cite{Baalrud_PRL_2013,Baalrud_POP_2019, Baalrud_POP_2014}. Here, Coulomb coupling strength is measured using the dimensionless parameter, $
\Gamma = (e^2/a)/ (k_B T)$, where $a = [3/(4\pi n)]^{1/3}$ is the Wigner-Seitz radius. 
A result of the generalized model is that there are several distinct Coulomb collision rates in strongly magnetized plasmas, which arise due to a dependence on the magnetic field in the binary collision process. 

This generalized kinetic theory was previously tested by computing the friction force on a test charge due to Coulomb collisions when moving through a strongly magnetized plasma, and comparing the results with first-principles molecular dynamics simulations~\cite{Jose_POP_2021}. Novel behaviors were predicted to arise from the asymmetries in the Coulomb collision rate caused by strong magnetization. For example, the friction force on a test charge was found to no longer be antiparallel to its velocity, but instead to have an additional transverse component in the plane of the velocity and magnetic field ~\cite{Lafleur_PPCF_2019,Lafleur_PPCF_2020,David_PRE_2020}. Later work showed that a third ``gyro-friction'' component in the Lorentz force direction also arises when the plasma is strongly coupled in addition to being strongly magnetized~\cite{Jose_POP_2021}. These perpendicular forces lead to non-intuitive features in the trajectory, such as an increase in the gyroradius of a fast test charge~\cite{Lafleur_PPCF_2019,Jose_POP_2021}. It has also been shown that strong magnetization influences macroscopic transport coefficients, such as electrical resistivity and conductivity~\cite{Baalrud_POP_2021,Jose_POP_2022}, and to cause a Barkas effect, where the transport properties depend on the sign of the charge of the interacting particles~\cite{Jose_POP_2022}. 

Other works calculating temperature relaxation rates for strongly magnetized plasmas employed perturbative methods~\cite{Silin_1963}, force correlation techniques~\cite{Kihara_JPSJ_1960,Kihara_RevModPhys_1960}, Fokker-Plank equations~\cite{Ichimaru_Phy_Fluids_1970}, linear response theory~\cite{Nersisyan_PRE_2011}, and binary collision models~\cite{Dong_POP_2013,Dong_POP_2013_2,Kihara_JPSJ_1959}. However, the inherent assumption of small-angle collisions in these theories limits them to the weakly coupled regime and to regimes where the gyroradius is larger than the distance of closest approach. Here, we  extend the parameter space to regimes where the gyroradus is smaller than the distance of the closest approach, and to regimes of strong Coulomb coupling (but with the limitation that $\Gamma_e \lesssim 20$).  
These extensions make the results applicable to experiments with strongly coupled components, such as ultra-cold neutral plasmas~\cite{Guthrie_POP_2021}, antimatter traps~\cite{Fajans_POP_2020}, and non-neutral plasmas~\cite{Beck_PRL_1992}. This work also extends the calculation of temperature relaxation to the extremely magnetized transport regime~\cite{Baalrud_PRE_2017} where the gyroradius is smaller than the distance of the closest approach. This is achieved by evaluating particle trajectories during interactions by numerically solving the equations of motion, instead of relying on a perturbative technique or weak interaction approximation.

Our previous work~\cite{Jose_POP_2023} calculating the ion-electron temperature relaxation rate found that when the plasma is strongly magnetized, parallel and perpendicular temperatures no longer relax at the same rate, which can lead to the development of temperature anisotropy. 
It was also found that the combination of oppositely charged interactions and strong magnetization caused the ion-electron relaxation rate in the parallel direction to be significantly suppressed, scaling inversely proportional to the magnetic field strength. However, the theory assumes that the ions do not gyrate during the collision event and that the electrons act as a heat bath. Here, we extend our previous work by relaxing these assumptions. 
The general electron-ion temperature relaxation problem is found to be complicated in this situation, involving 11 independent Coulomb collision rate coefficients. 
Each of these coefficients is evaluated from the kinetic theory. 
When focusing on a regime of strong electron magnetization ($\beta_e>1$) but weak ion magnetization ($\beta_i < 1$), the complexity is found to be reduced substantially, depending on just 3 collision rates: one associated with ion-ion collisions, one with electron-ion collisions in the parallel direction, and one with electron-electron temperature anisotropy relaxation. 

The outline of this paper is as follows. In Sec.~\ref{sec:theory}, a temperature evolution equation for ions and electrons is obtained. Section~\ref{sec:relaxation rates} discusses the relaxation rates for different magnetization strengths. Section~\ref{sec:discu} discusses the parallel and perpendicular temperature evolution of ions and electrons collisionaly relaxing from a non-equilibrium state to equilibrium. 

  \begin{figure} [!htb] 
\centerline{\includegraphics[width = 8cm]{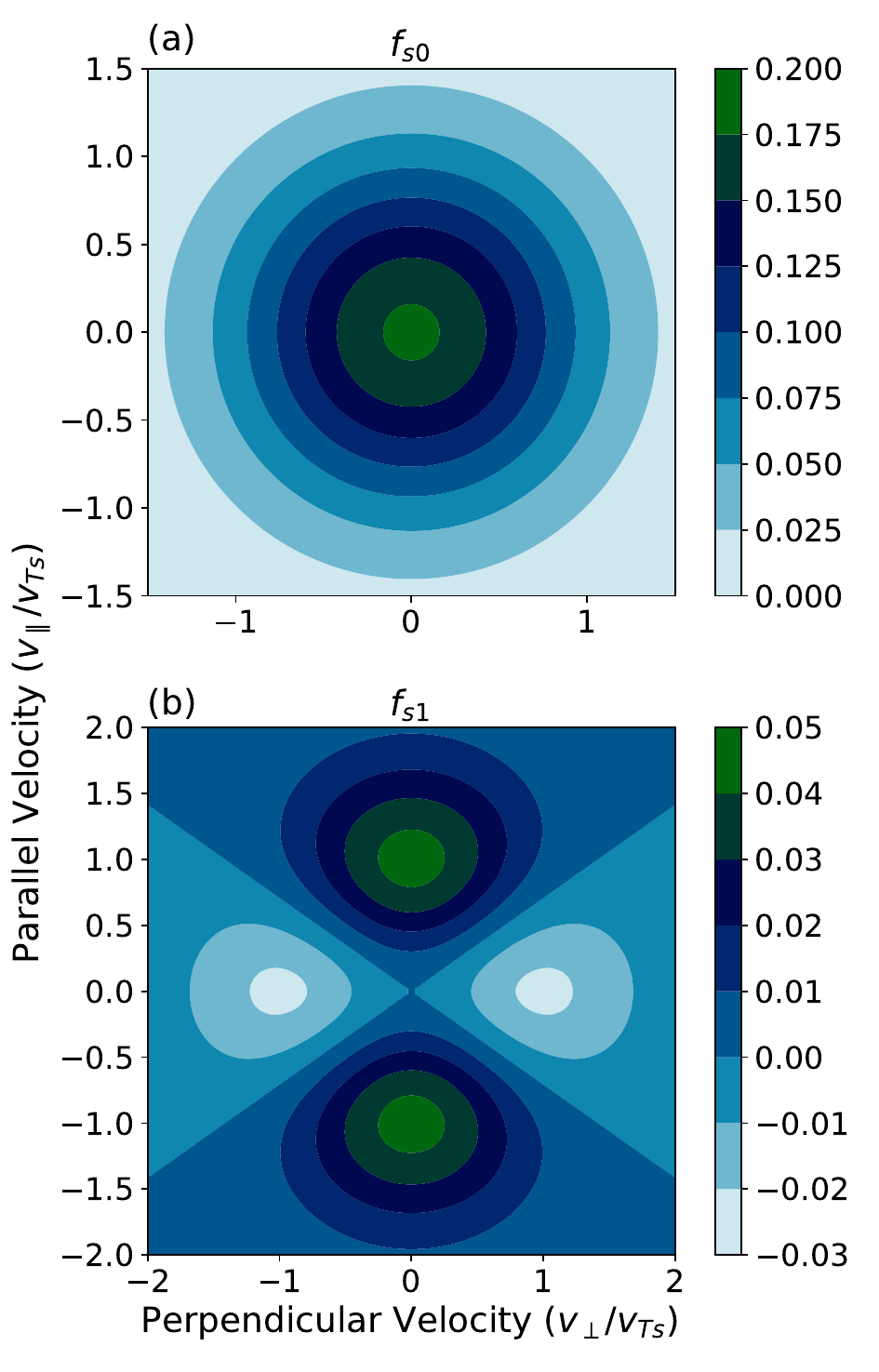}}
\caption {Illustration of the two terms in the expansion of the distribution function from Eq.~(\ref{eq:dist_exp}): (a) Maxwellian, and (b) first-order expansion in temperature anisotropy.}
  \label{fig:dists}
\end{figure}
 \begin{figure*} [!htb] 
\centerline{\includegraphics[width = 7.0in]{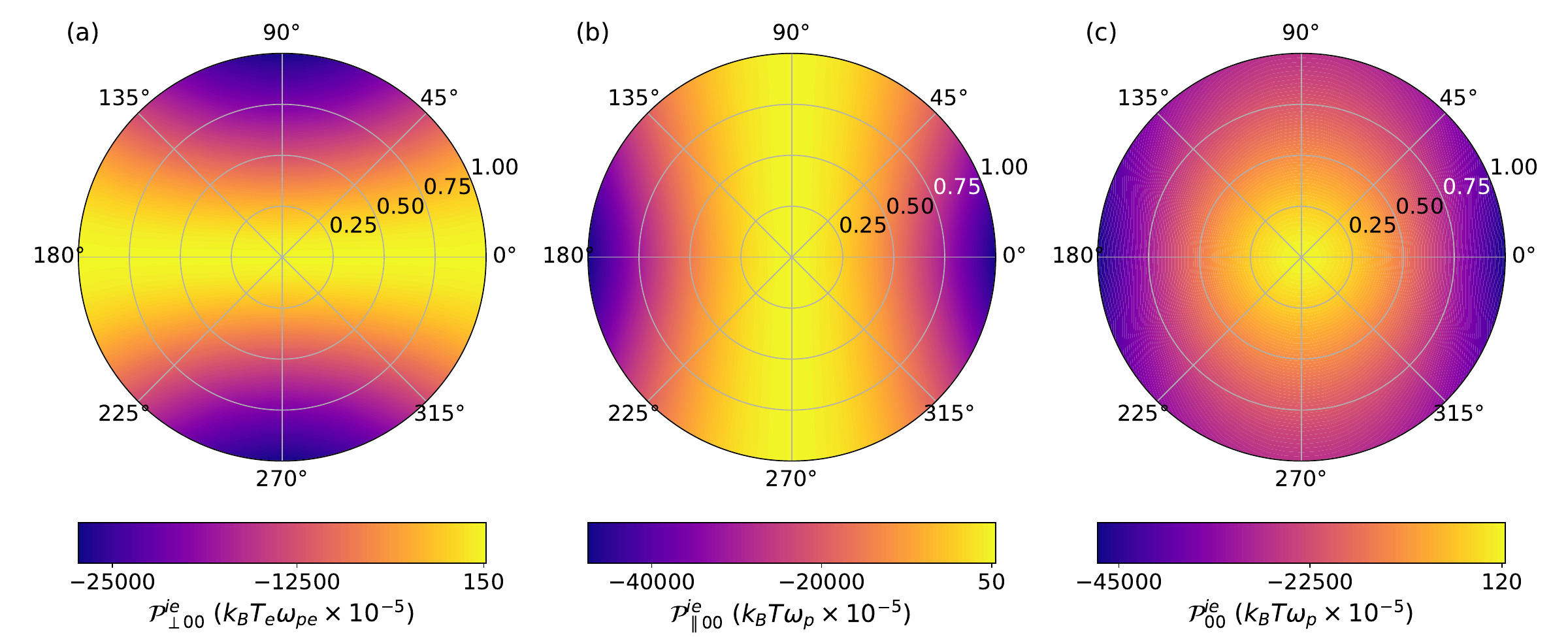}}
\caption {Polar plots of the energy exchange density components ($\mathcal{P}^{ie}_{\perp00}$ , $\mathcal{P}^{ie}_{\parallel00}$ and $\mathcal{P}^{ie}_{00}$) at $\Gamma_e= 1 $ and $\beta_e = 34$. The radial axis is the speed of the test charge ($v_0/v_{T_e}$) and the angle is the phase angle that the test charge velocity makes with the direction of the magnetic field ($\theta$).}
  \label{fig:polarqie00}
\end{figure*}

 \begin{figure*} [!htb] 
\centerline{\includegraphics[width = 7.0in]{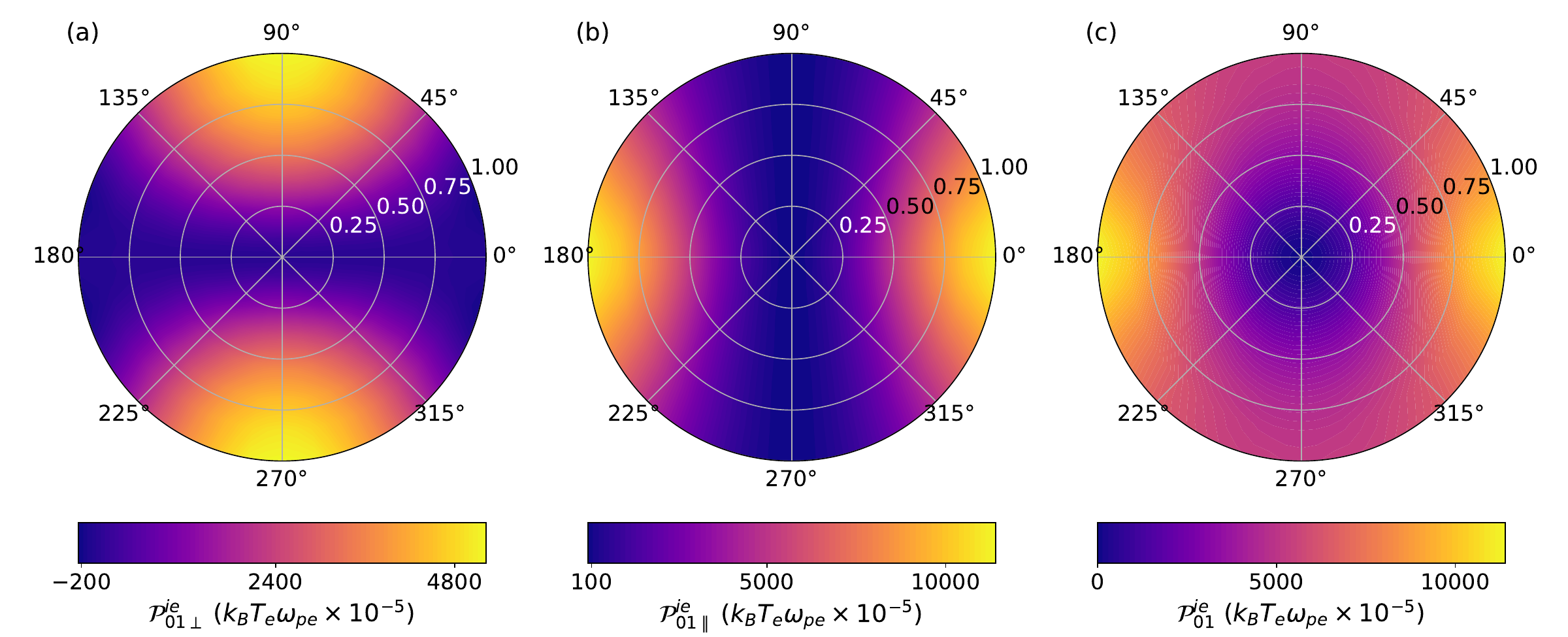}}
\caption {Polar plots of the energy exchange density components ($\mathcal{P}^{ie}_{\perp 01}$ , $\mathcal{P}^{ie}_{\parallel01}$ and $\mathcal{P}^{ie}_{01}$) at $\Gamma_e= 1 $ and $\beta_e = 34$. The radial axis is the speed of the test charge ($v_0/v_{T_e}$) and the angle is the phase angle that the test charge velocity makes with the direction of the magnetic field ($\theta$).}
  \label{fig:polarqie01}
\end{figure*}

\section{Theory\label{sec:theory}}

In weakly magnetized plasmas, the electron and ion temperatures are typically expected to remain isotropic as they relax. 
This is because the ion-ion and electron-electron energy relaxation rates are much faster than the electron-ion energy relaxation rate. 
Specifically, the temperature evolution equations have the form
\begin{subequations}
    \begin{eqnarray}
    \frac{d T_s}{d t} &=& -\mu^{s s^\prime} (T_s-T_{s^\prime})\\
    \frac{d T_{s^\prime}}{d t} &=& -\mu^{s^\prime s} (T_{s^\prime}-T_s) ,
\end{eqnarray}
\end{subequations}
where the standard energy relaxation rate due to Coulomb collisions is 
\begin{equation}
    \mu^{ss^\prime}_o = \frac{32 \sqrt{\pi} q_s^2 q_{s^\prime}^2 n_{s^\prime} \Xi_{ss^\prime}^{(1,1)}}{3 m_s m_{s^\prime} \bar{v}_{ss^\prime}^3} .
\end{equation}
Here, $s$ and $s^\prime$ label the two interacting species, $q_s$ are the charge states, $n_s$ the number density, $m_s$ the mass, and $\bar{v}_{ss^\prime}^2 = 2k_BT_s/m_s + 2k_BT_{s^\prime}/m_{s^\prime}$ an average thermal speed. 
In weakly coupled plasmas $\Xi_{ss^\prime}^{(1,1)} = \ln \Lambda$ is the standard Coulomb logarithm. 
Here, we use $\Xi_{ss^\prime}^{(1,1)}$ to denote the generalized Coulomb logarithm from mean force kinetic theory, as defined in \cite{Baalrud_PRL_2013}, to extend the results to strong coupling. 
For an electron-ion plasma, the mass scaling of the energy relaxation rate shows that the electron-electron relaxation is fastest ($\mu_o^{ee} \propto 1/\sqrt{m_e}$), ion-ion relaxation is the next fastest ($\mu_o^{ii} \propto \sqrt{m_e/m_i} \mu_o^{ee}$) and ion-electron relaxation is the slowest ($\mu_o^{ie} \propto \mu_o^{ei} \propto (m_e/m_i) \mu_o^{ee}$). 
This justifies assuming isotropic temperatures during the relaxation process. 

A different, and vastly more complicated, picture is expected with the plasma is strongly magnetized. 
Here, it is common to have different temperatures parallel and perpendicular to the magnetic field~\cite{Ott_PRE_2017}. 
This is possible because strong magnetization heavily suppresses the collisional anisotropy relaxation rates to a sufficient degree that the mass ratio dependence can be overcome and a temperature anisotropy sustained for a long time~\cite{Glinsky_Phys_Fluids_1992, Beck_PRL_1992}. Thus, the temperature evolution equations should account for temperature anisotropies ($ \Delta T_s = T_{s\parallel}- T_{s\perp}$) in addition to the temperature difference between species ($T_s - T_{s^\prime}$). 
For small deviations from equilibrium, the temperature evolution has a linear dependence on the temperature differences and anisotropies; i.e.,  $dT_s/dt \propto (T_s - T_{s^\prime}), \, \Delta T_s$ and $\Delta T_{s^\prime}$. Likewise, the temperature anisotropy evolutions is linearly proportional to the temperature differences and anisotropies; i.e.,  $d\Delta T_s/dt \propto (T_s - T_{s^\prime}), \, \Delta T_s$ and $\Delta T_{s^\prime}$. Thus, a general form of the temperature evolution has the form
\begin{subequations}\label{eq:genmus}
\begin{eqnarray}
    \frac{d T_s}{d t} &=& -\mu^{ss^\prime}_{00} (T_s-T_{s^\prime})+\mu^{s s^\prime}_{10} \Delta T_s + \mu^{s s^\prime}_{01} \Delta T_{s^\prime}, \\
    \frac{d \Delta T_s}{d t} &=& \mu^{ s s^\prime}_{A00} (T_s-T_{s^\prime})+\mu^{s s^\prime}_{A10} \Delta T_s+ \mu^{s s^\prime}_{A01} \Delta T_{s^\prime} \nonumber\\
    &&-\mu^{ss} \Delta T_s ,\\
        \frac{d T_{s^\prime}}{d t} &=& - \mu^{s^\prime s}_{00} (T_{s^\prime}-T_s)-  \mu^{s^\prime s}_{01} \Delta T_s- \mu^{s^\prime s}_{10} \Delta T_{s^\prime}, \\
    \frac{d \Delta T_{s^\prime}}{d t} &=&  \mu^{ s^\prime s}_{A00} (T_{s^\prime}-T_s)+  \mu^{s^\prime s}_{A01} \Delta T_s+  \mu^{s^\prime s}_{A10} \Delta T_{s^\prime} \nonumber\\
    &&-\mu^{s^\prime s^\prime} \Delta T_{s^\prime}.
\end{eqnarray}    
\end{subequations}  
Here, the relaxation rates, $\mu$, with superscript $ss^\prime$ corresponds to change in species $s$ due to collisions with $s^\prime$. 
The subscript $A$ indicates the evolution of temperature anisotropy and subscripts $0$ and $1$ corresponds to which species has the anisotropy. For example, $\mu^{s s^\prime}_{ij}$ corresponds to the relaxation rate between species $s$ with anisotropy of order $i$ and species $s^\prime$ with anisotropy of order $j$. 
The order of anisotropy is defined more specifically below, but 0 means no anisotropy (i.e., isotropic) and 1 means the linear order of anisotropy. 
Even though there are 14 different relaxation rates, the energy conservation ($n_s dT_s/dt = -n_{s^\prime} dT_{s^\prime}/dt$) constraint reduces the number of independent relaxation rates to 11. 
Specifically, this implies $n_s \mu^{ss^\prime}_{ij} = n_{s^\prime} \mu^{s^\prime s}_{ji}$, so $n_e \mu_{00}^{ei} = n_i \mu_{00}^{ie}$, $n_e \mu_{10}^{ei} = n_i \mu_{01}^{ie}$ and $n_e \mu_{01}^{ei} = n_i \mu_{10}^{ie}$.

The relaxation rates are obtained from the energy exchange density, which is the energy moment of the collision operator. The remainder of this section calculates the energy exchange densities of ions and electrons with particular attention given to the anisotropy in temperatures. 
The results are used to calculate the 11 temperature evolution coefficients in Eq.~(\ref{eq:genmus}) for conditions spanning weak to strong magnetization. 
It is also shown how temperature evolution equations simplify in the limit of weak magnetization. 


\subsection{Energy Exchange Density}

The temperature evolution equation can be obtained by taking an energy moment of the kinetic equation~\cite{baalrud2012transport, ichimaru2018statistical,nicholson1983introduction} 
\begin{equation}
    \frac{3}{2} n_s \frac{dT_s}{dt} = \sum_{s^\prime} Q^{ss^\prime}.
\end{equation}
Here, the collisional energy exchange rate between two species ($s$ and $s^\prime$) is given by the energy exchange density, which is the energy moment of the collision operator. Using the generalized Boltzmann collision operator, this is~\cite{Jose_POP_2023} 
\begin{equation}
\label{eq:Qs}
    \mathcal{Q}^{ss^\prime}=\int d^3 \vc{v} d^3 \vc{v}^{\prime} d\vc{s} |\vc{u}\cdot \vc{\hat{s}}| \frac{1}{2}m_s (\hat{v}^{ 2} - v^2) f_s (\vc{v}) f_{s^\prime} (\vc{v}^{\prime}) .
\end{equation}
Here, the surface integral is on the surface of the collision volume. The post collision velocity ($\vc{\hat{v}}$) is obtained by solving the equations of motion of the colliding particles inside the collision volume. The equations of motion of two charged particles interacting in the presence of a magnetic field is 
\begin{subequations}
    \label{eq:eoms}
\begin{eqnarray}
(m_s+m_{s^\prime})\frac{d\vc{V}}{dt} = m_{ss^\prime} \Big(\frac{\vc{u}}{c} \times \vc{B}\Big) \Big( \frac{q_s}{m_s}-\frac{q_{s^\prime}}{m_{s^\prime}} \Big) \nonumber \\
  + (q_s+q_{s^\prime}) \Big(\frac{\vc{V}}{c} \times \vc{B}\Big), \label{eoms1} \\
 m_{ss^\prime}\frac{d\vc{u}}{dt} = -\nabla \phi(r)+ m_{ss^\prime}^2 \Big(\frac{\vc{u}}{c} \times \vc{B}\Big) \Big(\frac{q_s}{m_s^2}+\frac{q_{s^\prime}}{m_{s^\prime}^2}\Big) \nonumber \\
 +m_{ss^\prime}  \Big(\frac{\vc{V}}{c} \times \vc{B}\Big) \Big(\frac{q_s}{m_s}-\frac{q_{s^\prime}}{m_{s^\prime}}\Big) .\label{eoms2}
\end{eqnarray}
\end{subequations}
where $\vc{V} = (m_s \vc{v}_s+m_{s^\prime} \vc{v}_{s^\prime})/(m_s+m_{s^\prime})$ is the center of mass velocity, $ m_{ss^\prime} = m_sm_{s^\prime}/(m_s+m_{s^\prime})$ is the reduced mass, and $\vc{u} = \vc{v}_s - \vc{v}_{s^\prime}$ is the relative velocity. The interaction between the particles is modeled using the potential of mean force, $\phi(r)$~\cite{Baalrud_PRL_2013,Baalrud_POP_2019}.  Here, the potential of mean force is obtained using the hypernetted-chain approximation for a one component plasma~\cite{hansen2013theory}. 
This applies to all interaction types (ion-ion, electron-ion and electron-electron) because all particles are taken to have the same charge ($q_s=e$). 
For a weakly coupled plasma, the potential of mean force is simply the Debye-H\"{u}ckel potential.

 \begin{figure*} [!htb] 
\centerline{\includegraphics[width = 7.0in]{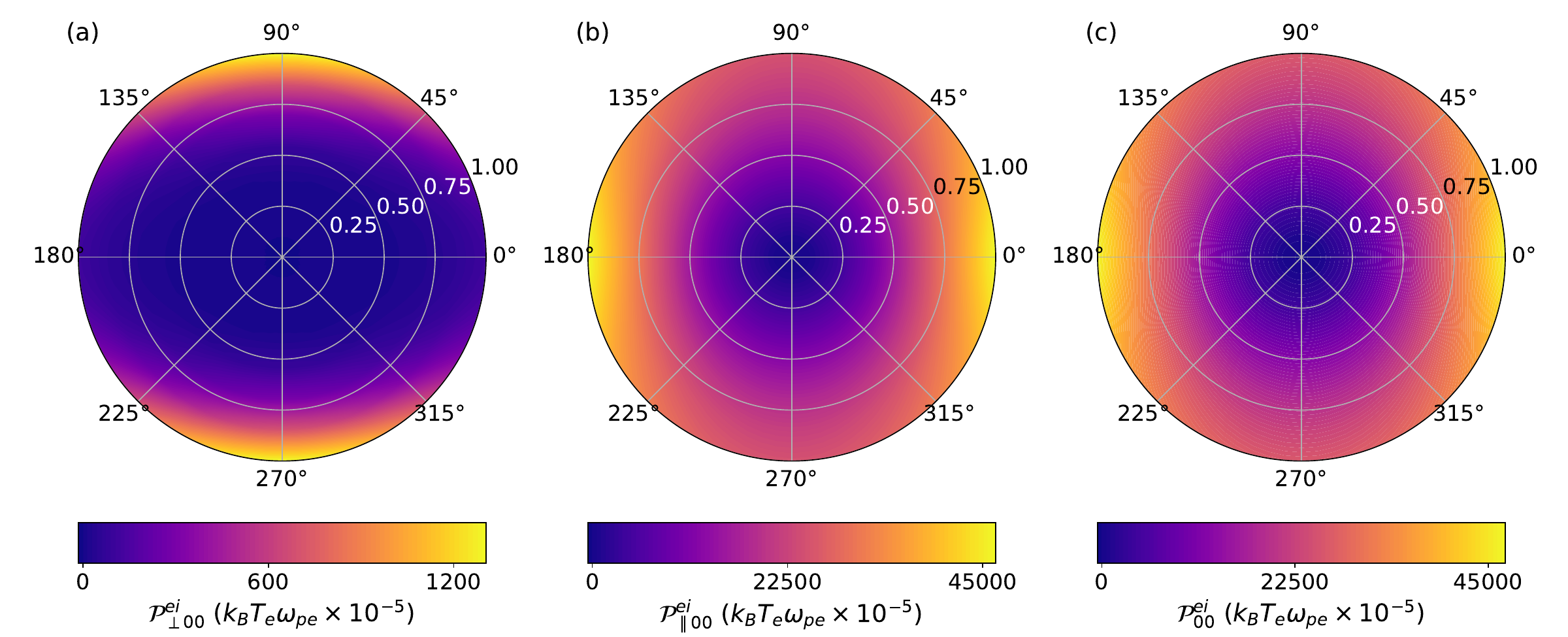}}
\caption {Polar plots of the energy exchange density components ($\mathcal{P}^{ei}_{\perp 00}$ , $\mathcal{P}^{ei}_{\parallel00}$ and $\mathcal{P}^{ei}_{00}$) at $\Gamma_e= 1 $ and $\beta_e = 17.2$. The radial axis is the speed of the test charge ($v_0/v_{T_e}$) and the angle is the phase angle that the test charge velocity makes with the direction of the magnetic field ($\theta$).}
  \label{fig:polarqei00}
\end{figure*}
 \begin{figure*} [!htb] 
\centerline{\includegraphics[width = 7.0in]{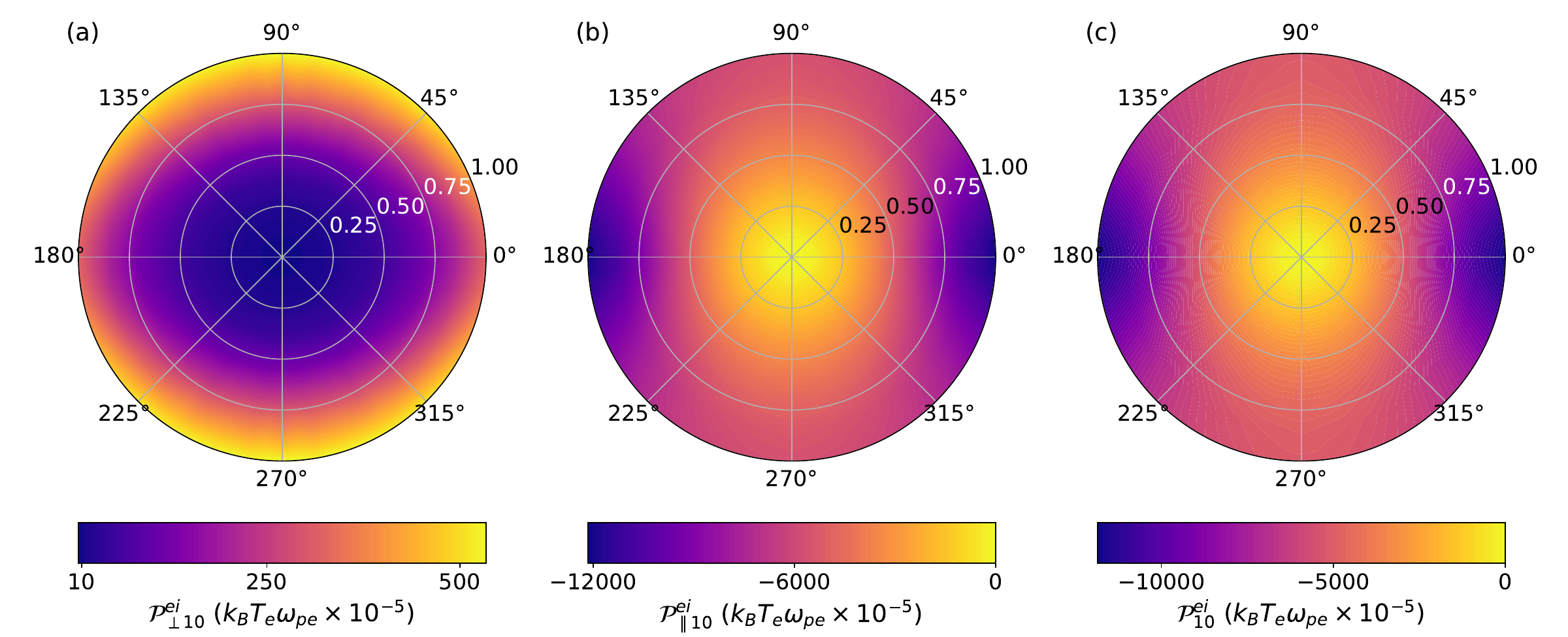}}
\caption {Polar plots of the energy exchange density components ($\mathcal{P}^{ei}_{\perp 10}$ , $\mathcal{P}^{ei}_{\parallel10}$ and $\mathcal{P}^{ei}_{10}$) at $\Gamma_e= 1 $ and $\beta_e = 17.2$. The radial axis is the speed of the test charge ($v_0/v_{T_e}$) and the angle is the phase angle that the test charge velocity makes with the direction of the magnetic field ($\theta$).}
  \label{fig:polarqei10}
\end{figure*}

Typically $f_s$ and $f_{s^\prime}$ are taken to be isotropic Maxwellian distributions with different temperatures. However, previous works~\cite{Jose_POP_2023,Glinsky_Phys_Fluids_1992, Ott_PRE_2017} have shown that strong magnetization breaks the symmetry of energy exchange leading to different temperature relaxation rates parallel and perpendicular to the magnetic field. For this reason, the distribution functions are modeled as anisotropic Maxwellians 
\begin{equation}
  f_s(\vc{v}) = \frac{n_s}{\pi ^{3/2} v_{Ts\parallel} v_{Ts\perp}^2} e^{-v_\perp^2/v_{Ts\perp}^2}e^{-v_\parallel^2/v_{Ts\parallel}^2},
\end{equation}
where $v_{Ts\perp}^2 = 2T_{s\perp}/m_s$, $v_{Ts\parallel}^2 = 2T_{s\parallel}/m_s$ and the parallel direction is along the magnetic field. 
Concentrating on the limit of small temperature anisotropy, the distribution functions are expanded to first order in $\Delta T_s = T_{s \parallel} -T_{s \perp}$, which provides
\begin{equation}
\label{eq:dist_exp}
    f_s=f_{s0}+f_{s1} \frac{\Delta T_s}{T_s}
\end{equation}
where
\begin{subequations}
\begin{eqnarray}
\label{eq:fs0}
    f_{s0} &=& \frac{n_s e^{-v^2/v_{Ts}^2}}{\pi ^{3/2}v_{Ts}^3}\\
    \label{eq:fs1}
    f_{s1} &=&  f_{s0} \frac{2v_\parallel^2-v_\perp^2}{3v_{Ts}^2} ,
\end{eqnarray}
\end{subequations}
and $v_{Ts}^2 = 2T_{s}/m_s$.
Note that $T_s = \frac{1}{3} T_{s\parallel} + \frac{2}{3} T_{s\perp}$, so $T_{s\parallel} = T_s + \frac{2}{3} \Delta T_s$ and $T_{s\perp} = T_s - \frac{1}{3} \Delta T_s$. Plots of the distribution functions are shown in Fig.~\ref{fig:dists}, illustrating that $f_{s1}$ peaks at the locations $(v_\parallel \simeq \pm v_{Ts}, v_\perp \simeq 0)$ and $(v_\parallel \simeq 0, v_\perp \simeq \pm v_{Ts})$.
Putting this expansion into Eq.~(\ref{eq:Qs}) shows that the energy exchange density can be split into three terms

\begin{equation} \label{Qtotssprime}
    \mathcal{Q}^{ss^{\prime}}  = \mathcal{Q}^{ss^{\prime}}_{00} +\mathcal{Q}^{ss^{\prime}}_{01} \frac{\Delta T_{s^\prime}}{T_{s^\prime}} +\mathcal{Q}^{ss^{\prime}}_{10}  \frac{\Delta T_{s}}{T_{s}}
\end{equation}
where,
\begin{subequations}
\begin{eqnarray}
&&\mathcal{Q}^{ss^{\prime}}_{00} = \int d^3 \vc{v} d^3 \vc{v}^{\prime} d\vc{s} |\vc{u}\cdot \vc{\hat{s}}| \frac{1}{2} m_s (\hat{v}^{ 2}-v^2) f_{s0} f_{s^\prime 0} \; \; \; \; \; 
 \label{eq: fs0fs0}\\
&&\mathcal{Q}^{ss^{\prime}}_{01} = \int d^3 \vc{v} d^3 \vc{v}^{\prime} d\vc{s} |\vc{u}\cdot \vc{\hat{s}}| \frac{1}{2} m_s (\hat{v}^{ 2}-v^2) \label{eq: fs0fs1}
f_{s0} f_{s^\prime 1} \\
&&\mathcal{Q}^{ss^{\prime}}_{10} = \int d^3 \vc{v} d^3 \vc{v}^{\prime} d\vc{s} |\vc{u}\cdot \vc{\hat{s}}| \frac{1}{2} m_s (\hat{v}^{ 2}-v^2)
f_{s1} f_{s^\prime 0} . \label{eq: fs1fs0}
\end{eqnarray}
\end{subequations}
Here, the subscript notation refers to the form of the distribution functions: ``0'' denotes the Maxwellian from Eq.~(\ref{eq:fs0}) and ``1'' the linear perturbation in anisotropy from Eq.~(\ref{eq:fs1}). Thus Eq.~(\ref{eq: fs0fs0}), accounts for the energy exchange when both species are Maxwellian, Eq.~(\ref{eq: fs0fs1}) is for the case when species $s$ is Maxwellian, and species $s^\prime$ has anisotropy, and Eq.~(\ref{eq: fs1fs0}) for the case when species $s$ has anisotropy and $s^\prime$ is Maxwellian.

Each of these terms can be split further into components parallel and perpendicular to the magnetic field. For example, $\mathcal{Q}^{ss^{\prime}}_{00} = \mathcal{Q}^{ss^{\prime}}_{\perp 00}+\mathcal{Q}^{ss^{\prime}}_{\parallel 00}$, where
\begin{subequations}
\begin{eqnarray}
 \mathcal{Q}^{ss^{\prime}}_{\perp 00} = \int d^3 \vc{v} d^3 \vc{v}^{\prime} d\vc{s} |\vc{u}\cdot \vc{\hat{s}}| \frac{1}{2} m_s (\hat{v}_{\perp}^{ 2}-v_{\perp}^2) 
f_{s0} f_{s^\prime 0}, \, \, \, \, \, \,\, \, \, \, \\
    \mathcal{Q}^{ss^{\prime}}_{\parallel 00} = \int d^3 \vc{v} d^3 \vc{v}^{\prime} d\vc{s} |\vc{u}\cdot \vc{\hat{s}}| \frac{1}{2} m_s (\hat{v}_{ \parallel }^{2}-v_{ \parallel }^2) 
f_{s0} f_{s^\prime 0}. \, \, \,\, \, \, \,\, \, \, \,
\end{eqnarray}
\end{subequations}
Analogous definitions apply for $\mathcal{Q}^{ss^{\prime}}_{01} = \mathcal{Q}^{ss^{\prime}}_{\perp 01}+\mathcal{Q}^{ss^{\prime}}_{\parallel 01}$ and $\mathcal{Q}^{ss^{\prime}}_{10} = \mathcal{Q}^{ss^{\prime}}_{\perp 10}+\mathcal{Q}^{ss^{\prime}}_{\parallel 10}$.
Since the distribution functions are written in terms of the temperature $T_s$ and the temperature anisotropy $\Delta T_s$, it is natural to write the evolution equation for these parameters rather than $T_{s\parallel}$ and $T_{s \perp}$.
The energy exchange density corresponding to evolution of total temperature, $T_s$ is $\mathcal{Q}^{ss\prime}$ and energy exchange density corresponding to evolution of temperature anisotropy, $\Delta T_s$ is $\mathcal{Q}^{ss^\prime}_A$,
where 
\begin{equation}
\label{eq:QA_def}
\mathcal{Q}^{ss^\prime}_{A}  \equiv  2 \mathcal{Q}^{ss^\prime}_\parallel - \mathcal{Q}^{ss^\prime}_\perp.
\end{equation}
As with $\mathcal{Q}^{ss^{\prime}}$, $\mathcal{Q}^{ss^{\prime}}_{A}$ can be split into different orders in the temperature anisotropy expansion,
\begin{eqnarray}\label{QAtotssprime}
    \mathcal{Q}^{ss^{\prime}}_{A}  = \mathcal{Q}^{ss^{\prime}}_{A00} + \mathcal{Q}^{ss^{\prime}}_{A01} \frac{\Delta T_{s^\prime}}{T_{s^\prime}} +\mathcal{Q}^{ss^{\prime}}_{A10}  \frac{\Delta T_{s}}{T_{s}}. 
\end{eqnarray}

The energy exchange density integrals are eight dimensional and solving them requires numerically solving the equations of motion, Eq.~(\ref{eq:eoms}), at each point in this eight dimensional space. 
Since this is computationally expensive, we break the integral into parts by first integrating over all the electron velocities and the collision surface as a function of ion velocity
\begin{equation}
\label{eq:Pie} 
\mathcal{P}^{ie} = \frac{1}{2} m_i \int   d^3v_e d\vc{s} |\vc{u}\cdot \vc{\hat{s}}|
(v_{i}^{\prime 2}-v_{i}^2)  f_e(\vc{v}_e),
\end{equation}
which is the energy exchanged by a single ion moving with a velocity $\vc{v}_0$ due to collisions with the electrons.  Similarly, the electron energy exchange density due to a collision with single ion can be expressed as
\begin{equation}
\label{eq:Pei}
\mathcal{P}^{ei} = \frac{1}{2} m_e \int  d^3v_e d\vc{s} |\vc{u}\cdot \vc{\hat{s}}|
(v_{e}^{\prime 2}-v_{e}^2)  f_e(\vc{v}_e).
\end{equation}
Though $\mathcal{P}^{ei}$ does not explicitly depend on the ion velocity~($\vc{v}_i$) in Eq.~(\ref{eq:Pei}), it implicitly depends on it because the change in electron energy in a collision with an ion is a function of the ion velocity.
The energy exchange density ($\mathcal{Q}$) of an ion distribution can be obtained by multiplying $\mathcal{P}$ by ion distribution function and integrating over all ion velocities, leading to
\begin{equation}
\label{eq:Qie2}
    \mathcal{Q}^{ie}  = \int d^3 v_i f_i(\vc{v}_i) \mathcal{P}^{ie} ,
\end{equation}
and
\begin{eqnarray}
\label{eq:Qei2}
 \mathcal{Q}^{ei} 
=\int d^3 v_i f_i(\vc{v}_i) \mathcal{P}^{ei}.
\end{eqnarray}
Similar to Eqs.~(\ref{eq:Qie2}) and~(\ref{eq:Qei2}), $\mathcal{Q}^{ie}_A$ and $\mathcal{Q}^{ei}_A$ can be obtained from the definition in Eq.~(\ref{eq:QA_def}) 
\begin{subequations}
\label{eq:Q_A}
\begin{eqnarray}
    \mathcal{Q}^{ie}_A  = \int d^3 v_0 f_i(\vc{v}_0) \mathcal{P}^{ie}_A ,\label{eq:QAie2} \\
     \mathcal{Q}^{ei}_A 
=\int d^3 v_0 f_i(\vc{v}_0) \mathcal{P}^{ei}_A, \label{eq:QAei2}
\end{eqnarray}
\end{subequations}
where
$\mathcal{P}^{ie}_A = 2\mathcal{P}^{ie}_\parallel - \mathcal{P}^{ie}_\perp$.


Numerical evaluation of the 5-D integrals to obtain $\mathcal{P}^{ie}$ and $\mathcal{P}^{ei}$ were performed similarly to our previous works~\cite{Jose_POP_2020,Jose_POP_2021,Jose_POP_2022, Jose_POP_2023, jose2023kinetic}. The equations of motion of the colliding particles, Eq.~(\ref{eq:eoms}), were solved using the ``DOP 853'' method~\cite{DOP853} and the energy moment integrals, Eqs.~(\ref{eq:Pie})~and~(\ref{eq:Pei}), were solved using the adaptive Monte Carlo integration code VEGAS~\cite{LEPAGE_2021_JCP,peter_lepage_2024_12687656}. The tolerance for the trajectory calculations was set to $10^{-8} - 10^{-10}$. The radius of the spherical collision volume was taken as $2.88a$, which was found in previous work~\cite{Jose_POP_2023} to be adequate  for the coupling strength of $\Gamma_e =1$ considered here.
This radius ensures that there is minimal interaction at the surface of the collision volume. The ratio of the mass of the ion ($m_i$) to the mass of the electrons ($m_e$) is set at 1000; $m_r=m_i/m_e =1000$. This particular mass ratio was chosen to correspond to previous and ongoing molecular dynamics simulations (MD)~\cite{David_PRE_2020}, which will be compared with the model in future work. The MD simulations become increasingly more expensive to run for higher mass ratios and magnetization strengths. For this mass ratio, ions becomes strongly magnetized ($\beta_i>1$), when electron magnetization, $\beta_e \gtrsim 32$ ($\beta_i/\beta_e = \sqrt{m_e/m_i}$).


 \begin{figure*} [!htb] 
\centerline{\includegraphics[width = 7.5in]{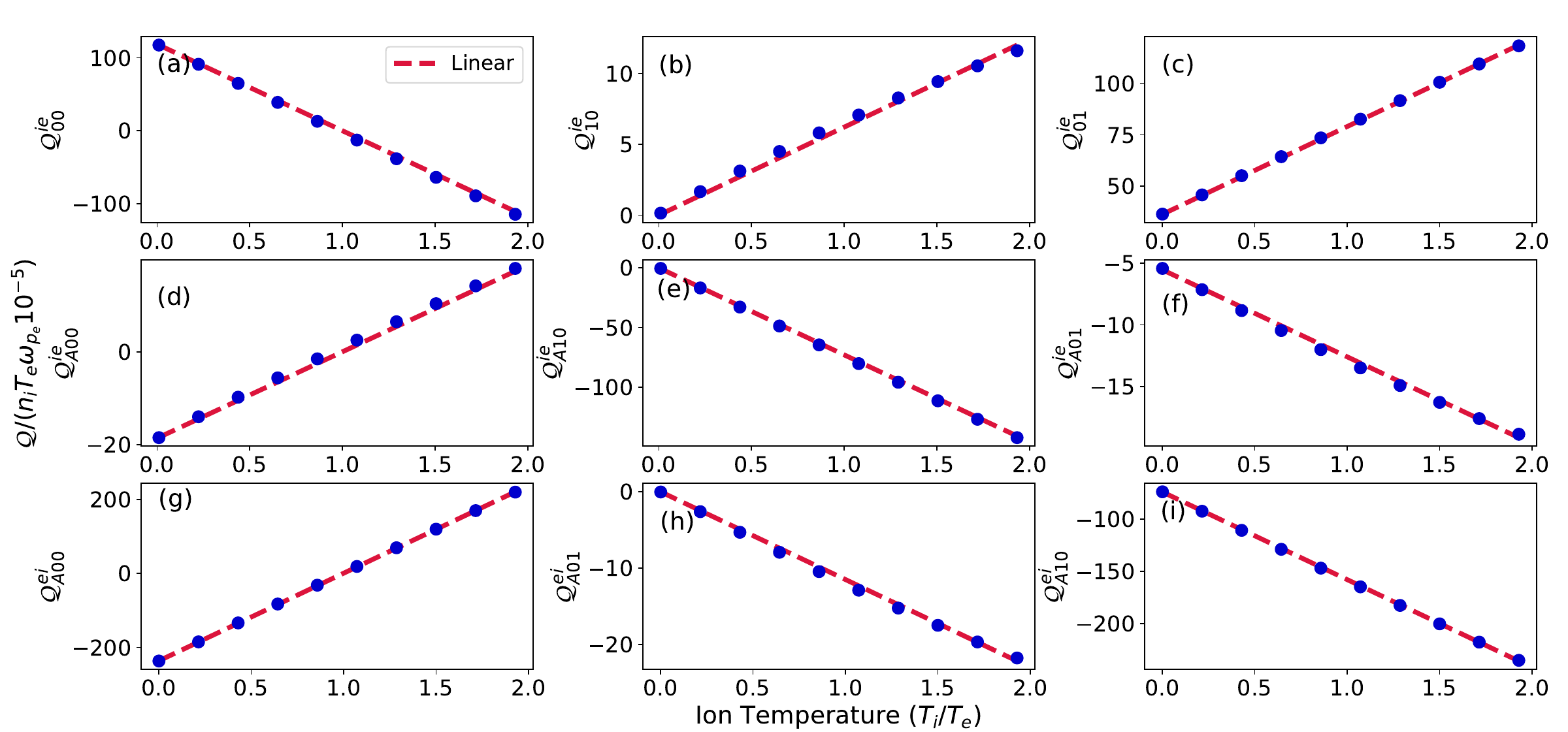}}
\caption {Energy exchange density of ions and electrons as a function of ion temperature. Circles denote solutions of Eqs. (\ref{eq:Qie2}), (\ref{eq:QAie2}) and (\ref{eq:QAei2}). The
dashed line shows the linear prediction from Eqs. (\ref{eq:linear ion first})-(\ref{eq:linear ion last}) and Eqs. (\ref{eq:linear electron first})-(\ref{eq:linear electron last}).  Here, the coupling strength is $\Gamma_e = 1$, and magnetization is $\beta_e = 34 $. }
  \label{fig:energy_densities_2}
\end{figure*}

\begin{table}
\centerline{\begin{tabular}{ |c |c |c| } 
 \hline
Magnetization ($\beta_e$)     & $\chi^{ie}_{01}$ & $\chi^{ie}_{A01}$ \\
 \hline
 2.2 & 3.69 & 0.89  \\ 
 4.4  & 2.23 & 1.43  \\
8.7& 1.52  & 1.78 \\
17.2  &  1.25 &  1.37  \\ 
34 &  1.17 &  1.28\\
66.9 &  1.10 &  1.04\\
131.8 &  0.97 &  0.268 \\
259.6 &  0.76 &  -0.74 \\
 \hline
\end{tabular}}
\caption{\label{tab: chiion} Fit values of $\chi^{ie}_{01}$($\beta$) and $\chi^{ie}_{A01}$($\beta$)for different magnetization strengths.}
\end{table}

\begin{table}
\centerline{\begin{tabular}{ |c |c | c| } 
 \hline
Magnetization ($\beta_e$)     & $\chi^{ei}_{A01}$ & $\chi^{ei}_{A10}$  \\
 \hline
 2.2 & 0.057   & 0.014  \\
 4.4  & 0.069 & 0.042  \\
8.7& 0.087  & 0.195 \\
17.2  &  0.097 &  0.868  \\ 
34  &  0.097 &  1.13  \\ 
 \hline
\end{tabular}}
\caption{\label{tab: chielectron} Fit values of $\chi^{ei}_{A01}$($\beta$) and $\chi^{ei}_{A10}$($\beta$)   for different magnetization strengths.}
\end{table}

The results of evaluating $\mathcal{P}^{ie}_{00}$, $\mathcal{P}^{ie}_{01}$, $\mathcal{P}^{ei}_{00}$ and $\mathcal{P}^{ei}_{10}$ are shown in Figs.~\ref{fig:polarqie00}-\ref{fig:polarqei10}. 
It is found that $\mathcal{P}^{ss^\prime}$ approaches a constant value in the limit that the test charge speed goes to zero, i.e., $v_o \rightarrow 0$, and is independent of the orientation of the test charge.
The energy exchange densities of ions are obtained from the polar plots of $\mathcal{P}^{ie}$ (Figs.~\ref{fig:polarqie00}-\ref{fig:polarqei10}) using Eqs.~(\ref{eq:Qie2}) and (\ref{eq:QAie2}). 
The electron energy exchange density components of the temperature anisotropy are obtained by evaluating Eq.~(\ref{eq:QAei2}). The results of these calculations are shown in Fig.~\ref{fig:energy_densities_2}, showing a linear scaling with respect to ion temperature.

In principle, evaluating the energy exchange densities requires evaluating the integral of each of these two-dimensional functions according to Eqs.~(\ref{eq:Qie2})-(\ref{eq:Q_A}) for each coefficient. 
However, since ions are much more massive than electrons, their velocity distribution is very narrow compared to that of the electrons. 
Thus, for the coefficients involving the product of Maxwellian ion and electron distribution functions ($Q_{00}^{ie}$, $Q_{00}^{ei}$, $Q_{A00}^{ie}$ and $Q_{A00}^{ei}$), we expect that ions can be modeled as a Dirac delta function for the purposes of computing the energy exchange densities in Eqs.~(\ref{eq:Qie2})-(\ref{eq:Q_A}). 
This corresponds to taking a test charge limit, as was done in previous work on the ion-electron temperature relaxation~\cite{Jose_POP_2023}. 
In these cases, the complexity of the computation is greatly reduced because three degrees of freedom are removed from the integrals and the energy exchange densities depend only on $\mathcal{P}$ in the limit of zero speed. 
Specifically, we define 
\begin{equation}
\mathcal{\tilde{P}}^{ei}  =  \lim_{\vc{v}_0 \to 0} \mathcal{P}^{ei} \ \textrm{and}\ \tilde{\mathcal{P}}^{ie} =   \lim_{\vc{v}_0 \to 0} \mathcal{P}^{ie} 
\end{equation}
as the limiting values corresponding to a test charge at zero velocity. 

Furthermore, in these cases we expect that the energy exchange density is proportional to the temperature difference: $Q_{00} \propto (T_i-T_e)$. 
In the limit of a small temperature difference, this can be shown from an expansion of $f_{e0}f_{i0}$ for $(T_i-T_e)/T \ll 1$. 
Figure~\ref{fig:energy_densities_2} shows from a direct evaluation of these coefficients that this linear relationship actually extends over a broad range of temperature differences. 
With this we find that the 00 coefficients have a relatively simple form 
\begin{subequations}
\label{eq:Q_00s}
\begin{eqnarray}
\mathcal{Q}^{ie}_{00} 
&=&n_i \mathcal{\tilde{P}}^{ie}_{00}\left(1-\frac{T_i}{T_e}\right) \label{eq:linear ion first}\\
\mathcal{Q}^{ie}_{A00} 
&=&n_i \mathcal{\tilde{P}}^{ie}_{A00}\left(1-\frac{T_i}{T_e}\right) \\
\mathcal{Q}^{ei}_{A00} 
&=&n_i  \mathcal{\tilde{P}}^{ei}_{A00} \left( 1-\frac{T_i}{T_e} \right) \label{eq:linear electron first}
\end{eqnarray}
\end{subequations}
and from energy conservation: $\mathcal{Q}^{ei}_{00} = -\mathcal{Q}^{ie}_{00}$. 
These expressions allow 4 of the 11 coefficients to be computed directly from just the low speed limit of $\mathcal{P}$. 


Coefficients associated with temperature anisotropy are more complicated to compute. 
Although the ion distribution function is narrow compared to the electron distribution function, $f_{s1} \rightarrow 0$ as $v\rightarrow 0$, as shown in figure~\ref{fig:dists}. 
Thus, the energy exchange rates cannot in general be computed from just the test particle limit. 
Although the coefficients involving a temperature anisotropy require evaluating the 8-dimensional integrals, it can still be useful to relate the results to the low velocity limit of $\mathcal{P}$. 
Based on direct evaluations, we observe that these coefficients are linear in $T_i/T_e$, but do not vanish when $T_e=T_i$ and have slopes that in general depend on $\beta_e$. 
This motivates a description of the energy exchange coefficients of the form
\begin{subequations}
\label{eq:Qlist}
\begin{eqnarray}
\mathcal{Q}^{ie}_{10} 
&=&-\frac{1}{3} n_i \mathcal{\tilde{P}}^{ie}_{A00}\frac{T_i}{T_e} \\
\mathcal{Q}^{ie}_{A10} 
&=&-\frac{1}{3} n_i \left( 2\mathcal{\tilde{P}}^{ie}_{00}+\mathcal{\tilde{P}}^{ie}_{A00} \right)\frac{T_i}{T_e} \\
 \mathcal{Q}^{ie}_{01} 
&=&n_i  \mathcal{\tilde{P}}^{ie}_{01} \left( 1+\chi^{ie}_{01} (\beta_e)\frac{T_i}{T_e} \right) \\
\mathcal{Q}^{ie}_{A01} 
&=&n_i  \mathcal{\tilde{P}}^{ie}_{A01} \left( 1+\chi^{ie}_{A01} (\beta_e)\frac{T_i}{T_e} \right)\label{eq:linear ion last} \\
\mathcal{Q}^{ei}_{A01} 
&=&n_i  \mathcal{\tilde{P}}^{ei}_{00} \chi^{ei}_{A01} (\beta_e) \left(\frac{T_i}{T_e} \right) \\
\mathcal{Q}^{ei}_{A10} 
&=&n_i  \mathcal{\tilde{P}}^{ei}_{A10} \left( 1+ \chi^{ei}_{A10} (\beta_e) \frac{T_i}{T_e} \right). \label{eq:linear electron last}
\end{eqnarray}
\end{subequations}
and where energy conservation provides $\mathcal{Q}^{ei}_{01} = -\mathcal{Q}^{ie}_{10}$ and $\mathcal{Q}^{ei}_{10} = -\mathcal{Q}^{ie}_{01}$

Although computing the factors $\chi$ in Eq.~(\ref{eq:Qlist}) requires the full integration of the polar plots for $\mathcal{P}$, casting them in terms of $\tilde{\mathcal{P}}$ can be useful because in some limiting cases $\tilde{\mathcal{P}}$ can get very small, justifying the smallness of these terms without necessarily evaluating $\chi$. 
Finally, we note that the main computational challenge is the four $\chi$ factors, which depend on the magnetization strength. Values of these coefficients computed for different magnetization strengths are provided in tables \ref{tab: chiion} and \ref{tab: chielectron}.
These show that the values are in general of order unity over a broad range of magnetization strengths, but that there is still some variation with magnetic field strength, including the possibility of a change in sign.

\subsection{Temperature Evolution}

As described above, the temperature evolution equations obtained from the moments of the generalized Boltzmann equation are related to the energy exchange densities by
\begin{subequations}
\label{eq:Ts}
\begin{eqnarray}
\frac{3 n_s}{2} \frac{dT_s}{dt} &=& \mathcal{Q}^{ss^\prime}, \label{Tev}\\
n_s \frac{d \Delta T_s}{dt} &=&\mathcal{Q}^{ss^\prime}_A +\mathcal{Q}^{ss}_A, \label{DTev}
\end{eqnarray}
\end{subequations}
where $\mathcal{Q}^{ss}_A$ is the energy exchange density due to intra-species collisions. The previous subsection only considered the contribution of ion-electron and electron-ion collisions. However, when a temperature anisotropy is present contributions from ion-ion and electron-electron collisions also need to be included as they act to relax the anisotropy. 
Analogously to the interspecies collisions derived above, the energy exchange density due to intraspecies collisions can be written as
\begin{subequations}
\label{eq:Qss}
\begin{eqnarray}
    \mathcal{Q}^{ii}_A &=& -3 \nu^{ii} \Delta T_i \\
    \mathcal{Q}^{ee}_A &=& -3 \nu^{ee} \Delta T_e. 
\end{eqnarray}
\end{subequations}
Here, $\nu^{ii} $ and $\nu^{ee} $ are the collisional equipartition rates of ions and electrons. Previous works have shown that the temperature anisotropy relaxation is severely suppressed when the plasma is strongly magnetized~\cite{Glinsky_Phys_Fluids_1992,Beck_PRL_1992}.
For example, O'Neil's Boltzmann-like kinetic theory for weakly coupled strongly magnetized one-component plasma has been applied to show an rapid decrease in the temperature anisotropy relaxation rate that agreed well with experiments in non-neutral plasmas. 
Here, the calculations of the generalized Boltzmann kinetic equation extend these results to the transition regime between weak and strong magnetization, as well as to the strongly coupled regime.

Putting the definitions from Eqs.~(\ref{eq:Qlist}) and (\ref{eq:Qss}) into Eq.~(\ref{eq:Ts}) provides the full set of electron-ion temperature evolution equations
\begin{widetext}
\begin{subequations}
\label{eq:rates_1}
\begin{eqnarray}
    \frac{d T_i}{d t} &=& -\nu^{ie}_{00} (T_i-T_e)+\nu^{ie}_{10} \Delta T_i+ \nu^{ie}_{01} \left( 1+\chi^{ie}_{01}  \frac{T_i}{T_e} \right) \Delta T_e 
    \label{Ti_Eqn} \\
    \frac{d \Delta T_i}{d t} &=& \nu^{ ie}_{A00} (T_i-T_e)-\nu^{ie}_{A10} \Delta T_i- \nu^{ie}_{A01} \left( 1+\chi^{ie}_{A01}  \frac{T_i}{T_e} \right) \Delta T_e   -3 \nu^{ii} \Delta T_i 
    \label{eq:DTi_eq}\\
    \frac{d T_e}{d t} &=& -\frac{n_i}{n_e}\nu^{ei}_{00} (T_e-T_i)-\frac{n_i}{n_e} \nu^{ei}_{01} \Delta T_i-\frac{n_i}{n_e}\nu^{ei}_{10} \left( 1+\chi^{ei}_{10}  \frac{T_i}{T_e} \right) \Delta T_e 
    \label{eq:Te_eq}\\
    \frac{d \Delta T_e}{d t} &=& \frac{n_i}{n_e}\nu^{ ei}_{A00} (T_e-T_i)+ \frac{n_i}{n_e}\nu^{ei}_{A01} \chi^{ei}_{A01} \Delta T_i+\frac{n_i}{n_e}\nu^{ei}_{A10} \left( 1+ \chi^{ei}_{A10}   \frac{T_i}{T_e} \right) \Delta T_e  -3 \nu^{ee} \Delta T_e
    \label{DTe_Eqn}
\end{eqnarray}
\end{subequations}
where $\chi^{ei}_{10} = \chi^{ie}_{01}$ and the coefficients $\nu$ are defined from the energy exchange densities of a test charge ($\tilde{P}$) as described in the list in Appendix~\ref{app:rates}. 
Here, the rate coefficients ($\mu$) are written in a way to explicitly show the contributions from the test charge limit part ($\nu$) and the ion distribution part that cannot be described as a Dirac delta function ($\chi$). 
They can equivalently be written in the form of Eq.~(\ref{eq:genmus}), where the rate coefficients ($\mu$) are related to the $\nu$ and $\chi$ coefficients as provided in Appendix~\ref{app:rates}.

Writing the temperature evolution equation in terms of the total temperature, $T_s$ and anisotropy, $\Delta T_s$ exploits the energy conservation and thus decreases the number of relaxation rates required to compute. However, writing these equations in terms of evolution of $T_{s\parallel}$ and $T_{s\perp}$ explicitly shows the relaxation rates parallel and perpendicular to the magnetic field. The resulting temperature evolution equation in terms of  $T_{s\parallel}$ and $T_{s\perp}$ are
    \begin{subequations}
        \label{eq: rates_3}
\begin{eqnarray}
    \frac{d T_{i\parallel}}{d t} &=& -\nu^{ie}_{\parallel 00} (T_i-T_e)+\nu^{ie}_{\parallel 10} \Delta T_i+ \nu^{ie}_{\parallel 01}  \left( 1+\chi^{ie}_{\parallel 01}  \frac{T_i}{T_e} \right) \Delta T_e -2\nu^{ii} \Delta T_i\\
    \frac{d T_{i\perp}}{d t} &=& -\nu^{ie}_{\perp 00} (T_i-T_e)+\nu^{ie}_{\perp 10} \Delta T_i+ \nu^{ie}_{\perp 01}  \left( 1+\chi^{ie}_{\perp 01}  \frac{T_i}{T_e} \right) \Delta T_e +\nu^{ii} \Delta T_i\\
    \frac{d T_{e\parallel}}{d t} &=& -\frac{n_i}{n_e}\nu^{ei}_{\parallel 00} (T_e-T_i)-\frac{n_i}{n_e} \nu^{ei}_{\parallel 01} \chi^{ei}_{\parallel 01} \Delta T_i
    -\frac{n_i}{n_e} \nu^{ei}_{\parallel 10} \left( 1+\chi^{ei}_{\parallel 10}  \frac{T_i}{T_e} \right) \Delta T_e -2\nu^{ee} \Delta T_e \\
    \frac{d T_{e \perp}}{d t} &=& -\frac{n_i}{n_e}\nu^{ei}_{\perp 00} (T_e-T_i)-\frac{n_i}{n_e} \nu^{ei}_{\perp 01} \chi^{ei}_{\perp 01}  \Delta T_i-\frac{n_i}{n_e}  \nu^{ei}_{\perp 10} \left( 1+\chi^{ei}_{\perp10}  \frac{T_i}{T_e} \right)  \Delta T_e +\nu^{ee} \Delta T_e
    \label{dTe_perpdt}
\end{eqnarray}
    \end{subequations}
\end{widetext}
where the rate coefficients definitions are 
\begin{subequations}
\label{eq:rate_ptoA}
\begin{eqnarray}
    \nu_{\parallel ij}^{ss^\prime} =  \nu_{ij}^{ss^\prime} - \frac{2}{3} \nu_{Aij}^{ss^\prime}, \\
        \nu_{\perp ij}^{ss^\prime} = \nu_{ij}^{ss^\prime} + \frac{1}{3} \nu_{Aij}^{ss^\prime}.
\end{eqnarray}
\end{subequations}
and $\chi^{ie}_{\parallel 01} $, $\chi^{ie}_{\perp 01} $,$\chi^{ei}_{\parallel 10}$, $\chi^{ei}_{\perp 10}$ , $\chi^{ei}_{\parallel 01}$ and $\chi^{ei}_{\perp 01}$ are determined from
\begin{subequations}
\begin{eqnarray}
        \chi^{ss^\prime}_{\parallel ij} = \frac{\nu^{ss^\prime}_{ij} \chi^{ss^\prime}_{ij}-\frac{2}{3} \nu^{ss^\prime}_{Aij} \chi^{ss\prime}_{Aij}}{\nu^{ss^\prime}_{ij} -\frac{2}{3} \nu^{ss^\prime}_{Aij} }, \\
    \chi^{ss^\prime}_{\perp ij} = \frac{\nu^{ss^\prime}_{ij} \chi^{ss^\prime}_{ij}+\frac{1}{3} \nu^{ss^\prime}_{Aij} \chi^{ss^\prime}_{Aij}}{\nu^{ss^\prime}_{ij} +\frac{1}{3} \nu^{ss^\prime}_{Aij} } .
    \label{eq:chi_perp}
\end{eqnarray}
\end{subequations}
Here, $\chi^{ei}_{01}=1$. Since  $\nu^{ie}_{10}$ and $\nu^{ie}_{A10}$ can be written in terms of $\nu^{ie}_{00}$ and , $\nu^{ie}_{A00}$ [Eqs.~(\ref{apnuie10}) and (\ref{apnuaie10})], we obtain
$\nu_{\parallel 10}^{ie} = -\frac{2}{3} \nu_{\parallel 00}^{ie}$ and $\nu_{\perp 10}^{ie} = \frac{1}{3} \nu_{\perp 00}^{ie}$.

\subsection{Weakly magnetized limit}\label{subsection:weak}

Analytical expressions for the the energy exchange densities can be obtained in the weakly magnetized limit ($\beta_e \ll 1$). 
These calculations are shown in Appendix~\ref{sec:weak_results}. 
Because the collisions are not influenced by the magnetic field in this limit, all the terms that have to do with asymmetry in the collision process vanish: $\mathcal{Q}^{ie}_{10} = \mathcal{Q}^{ie}_{01} = \mathcal{Q}^{ei}_{10} = \mathcal{Q}^{ei}_{01} = \mathcal{Q}^{ie}_{A00} =\mathcal{Q}^{ei}_{A00} =0$, implying that $ \nu^{ie}_{10} = \nu^{ie}_{01} = \nu^{ei}_{10} = \nu^{ei}_{01} = \nu^{ie}_{A00} = \nu^{ei}_{A00} = 0$.
The temperature evolution equations in this case reduce to
\begin{subequations}
\begin{eqnarray}
    \frac{d T_i}{d t} &=& -\nu^{ie}_{00} (T_i-T_e)\\
    \frac{d \Delta T_i}{d t} &=& -\nu^{ie}_{A10} \Delta T_i-\nu^{ie}_{A01} \left( 1+\chi^{ie}_{A01}  \frac{T_i}{T_e} \right) \Delta T_e   \nonumber \\
    && -3 \nu^{ii} \Delta T_i\\
    \frac{d T_e}{d t} &=& -\frac{n_i}{n_e}\nu^{ie}_{00} (T_e-T_i) \\
    \frac{d \Delta T_e}{d t} &=&  \frac{n_i}{n_e}\nu^{ei}_{A01} \chi^{ei}_{A01} \Delta T_i+\frac{n_i}{n_e}\nu^{ei}_{A10} \left( 1+ \chi^{ei}_{A10}   \frac{T_i}{T_e} \right) \Delta T_e  \nonumber \\
    && -3 \nu^{ee} \Delta T_e
\end{eqnarray}
\end{subequations}
where the relaxation rates in units of electron plasma frequency are
\begin{subequations}
\label{eq:weakB_rates}
\begin{eqnarray}
    \frac{\nu^{ie}_{00}}{\omega_{pe}} &=& \frac{\nu^{ei}_{00}}{\omega_{pe}} = \frac{\nu^{ie}_{A10}}{\omega_{pe}} = - \frac{2}{3} \frac{\nu^{ei}_{A01} }{\omega_{pe}} = \frac{\sqrt{8} \Gamma_e^{3/2} \Xi ^{(1,1)}}{\sqrt{3 \pi } m_r} \\
    \frac{\nu^{ie}_{A01}}{\omega_{pe}} &=& \frac{-\sqrt{8} \Gamma_e^{3/2} (4\Xi ^{(1,2)}-3\Xi ^{(2,2)})}{10 \sqrt{3 \pi } m_r} \\
    \frac{\nu^{ee}}{\omega_{pe}} &=& - \frac{1}{3\sqrt{2}} \frac{\nu^{ie}_{A01}}{\omega_{pe}} = \frac{ \Gamma_e^{3/2} \Xi ^{(2,2)}}{5 \sqrt{3 \pi }}\\
\frac{\nu^{ii}}{\omega_{pe}}  &=& \sqrt{\frac{n_i}{n_e}}\frac{ \Gamma_i^{3/2} \Xi ^{(2,2)}}{5 \sqrt{3 \pi }\sqrt{m_r}}
\end{eqnarray}
\end{subequations}
and
\begin{subequations}
\begin{eqnarray}
    \chi^{ie}_{A01} &=& \frac{10 \Xi^{(1,1)} -4 \Xi^{(1,2)} }{4 \Xi^{(1,2)} -3 \Xi^{(2,2)} } \\
    \chi^{ei}_{A01} &=& \frac{-10 \Xi^{(1,1)} +3 \Xi^{(2,2)} }{15 \Xi^{(1,1)} } \\
    \chi^{ei}_{A10} &=&  \mathcal{O} \left(\frac{1}{m_r}\right).
\end{eqnarray}
\end{subequations}
In these expressions, $m_r=m_i/m_e$ is the ion-to-electron mass ratio, and the generalized Coulomb logarithm
\begin{eqnarray}
    \Xi^{(l,k)} = \frac{1}{2}\int d \xi \xi^{2k+3} e^{-\xi^2} \sigma^{(l)}/\sigma_0
\end{eqnarray}
extends the traditional plasma theory into the strongly coupled regime ($\Gamma \lesssim 20$)~\cite{Baalrud_PRL_2013, Baalrud_POP_2014}. Here, $\sigma_0 \equiv \pi e^4 /(m_{ss^\prime}^2\bar{v}_{ss^\prime}^2)$, is a reference cross section, $\xi^2 \equiv u^2/\bar{v}_{ss^\prime}^2$, and $\sigma^{(l)}$ is the $l$\textsuperscript{th} momentum scattering cross section obtained using the potential of mean force [Eq. (10) in Ref.~\onlinecite{Baalrud_POP_2014}]. 

Due to the mass ratio scaling, ion-ion and electron-electron interactions occur at a faster rate than electron-ion interactions; by a factor of $\sqrt{m_r}$ and $m_r$, respectively. In addition to the intra-species collisions, electron anisotropy relaxation through collisions with ions ($\nu^{Aei}_{10}$ ) is independent of mass ratio and is of same order of magnitude as electron-electron collisions. 
Thus, in weakly magnetized plasmas, any temperature anisotropy formed is quickly relaxed. 
Concentrating on the timescale of electron-ion temperature relaxation, the evolution of temperature anisotropy can be neglected, which leads to the common evolution equations 
\begin{eqnarray}
    \frac{d T_i}{d t} &=& -\nu^{ie}_{00} (T_i-T_e)\\
    \frac{d T_e}{d t} &=& -\frac{n_i}{n_e}\nu^{ei}_{00} (T_e-T_i) .
\end{eqnarray}
Thus, the quick intra-species anisotropy relaxation compared to that of inter-species relaxation greatly simplifies the temperature relaxation process in the weakly magnetized limit. 
This simplification is not possible at strong magnetization because perpendicular energy exchange is strongly suppressed, and thus, the above mass ratio argument for neglecting the temperature anisotropy is invalid. Accounting for the evolution of temperature anisotropy makes the relaxation process more complex.

\section{Relaxation rates \label{sec:relaxation rates}}

\begin{figure*} [!htb] 
\centerline{\includegraphics[width = 1.80\columnwidth]{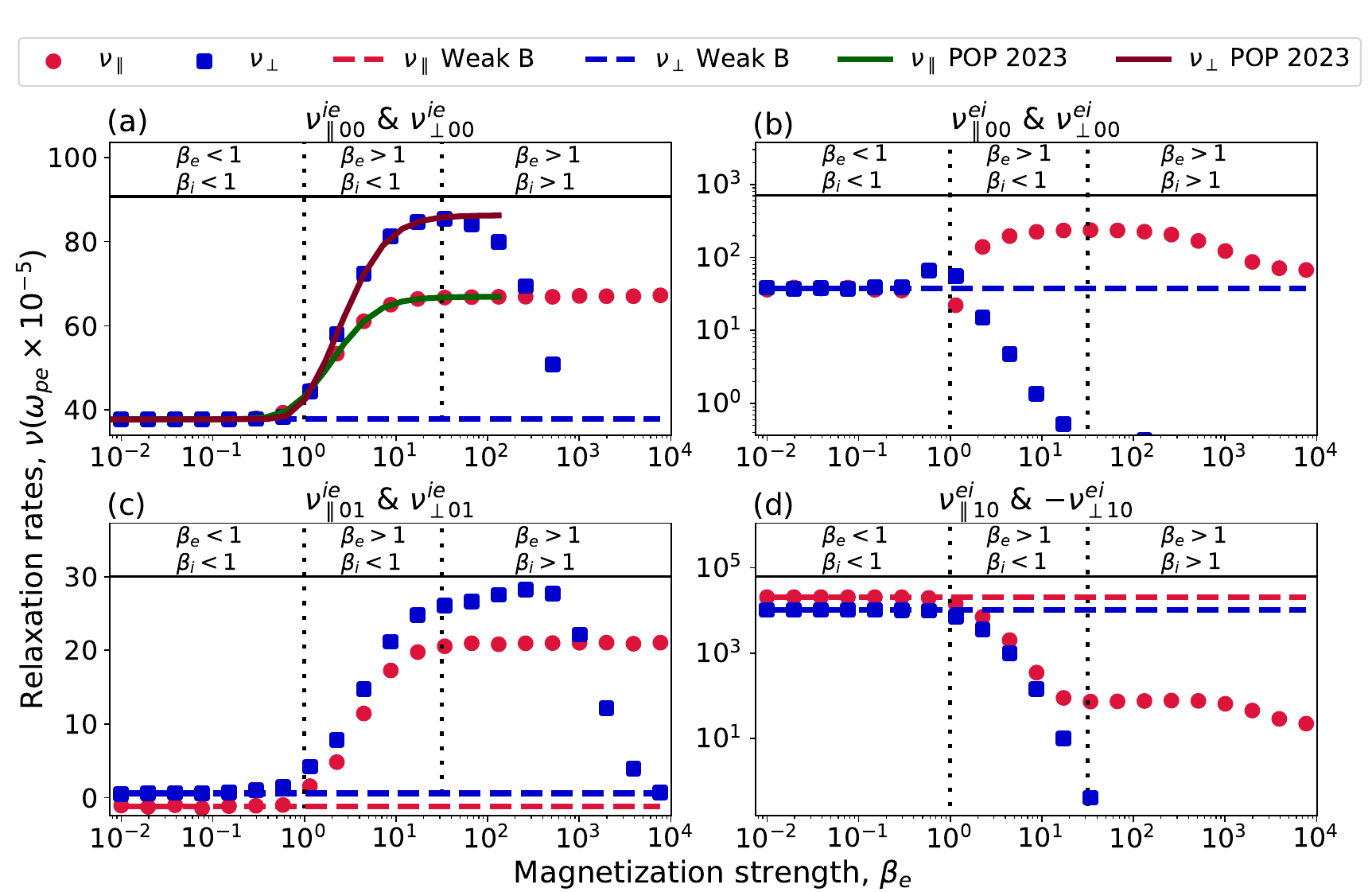}}
\caption {Electron and ion temperature relaxation rates as a function of electron magnetization strength ($\beta_e$). Here, the electron coupling strength is $\Gamma_e =1$. Results in the weak magnetization limit are computed from subsection~\ref{subsection:weak}.}
  \label{fig:relaxrates}
\end{figure*}

Results of the relaxation rate coefficients as a function of magnetization strength are shown in Fig.~\ref{fig:relaxrates}. 
Panel (a) shows the coefficients associated with the energy relaxation of ions due to collisions with electrons when both species have Maxwellian distributions, $\nu^{ie}_{\parallel 00}$ and $\nu^{ie}_{\perp 00}$. 
Solid lines show the results from our previous work~\cite{Jose_POP_2023}, which assumed that the ions are not strongly magnetized and therefore do not gyrate during collisions. As in that previous work, the parallel and perpendicular relaxation rates are equal in the weakly magnetized regime ($\beta_e <1$). This implies that these terms do not lead to the formation of a temperature anisotropy in this limit. 
In contrast, the parallel and perpendicular relaxation rates increase in magnitude when the plasma becomes strongly magnetized ($\beta_e > 1)$, but at different rates, implying that this collision process acts to form a temperature anisotropy.

In the transition to strong electron magnetization, the new results agree with the previous calculations where the ions were treated as unmagnetized. 
However, at a sufficiently large magnetization strength ($\beta_e \gtrsim 100$), the perpendicular relaxation rate decreases rapidly in the new calculation. 
This corresponds to a regime where both electrons and ions are strongly magnetized ($\beta_i \gtrsim 1$), and therefore was not captured in the previous work~\cite{Jose_POP_2023}. 
The rapid decrease of the perpendicular relaxation rate to near zero implies that at high magnetization the perpendicular kinetic energy becomes an adiabatic invariant. 
This is similar to what has been observed in one-component plasmas~\cite{Oneil_Phys_Fluids_1983,Oneil_Phys_Fluids_1985,Glinsky_Phys_Fluids_1992}. 
In contrast, the parallel relaxation rate remains at the same value whether the ions are strongly magnetized or not. 
This implies that the parallel energy relaxation rate of ions is influenced by the electron gyromotion, but is independent of the ion gyromotion.

Figure~\ref{fig:relaxrates}(b) shows the coefficients associated with energy relaxation of electrons due to collisions with ions when both species have Maxwellian distributions, $\nu^{ei}_{\parallel 00}$ and $\nu^{ei}_{\perp 00}$. As expected, the perpendicular and parallel relaxation rates are equal in the weakly magnetized regime ($\beta_e <1$). 
They are also equal to the corresponding ion-electron relaxation rates from Fig.~\ref{fig:relaxrates}(a): $\nu_{\parallel 00}^{ie} = \nu_{\perp 00}^{ie} = \nu_{\parallel 00}^{ei} = \nu_{\perp 00}^{ei}$. 
These results imply that the electron and ion temperatures will relax symmetrically (without anisotropy) and at equal rates, as expected when the magnetic field does not influence the relaxation process. 
In contrast, both of these symmetries are broken in the strongly magnetized regime ($\beta_e > 1$). 
Specifically, not only are the parallel and perpendicular rates unequal, but the parallel rate for ion-electron energy relaxation is unequal to the parallel rate for electron-ion energy relaxation ($\nu_{\parallel 00}^{ie} \neq \nu_{\parallel 00}^{ei}$) and similarly in the perpendicular direction ($\nu_{\perp 00}^{ie} \neq \nu_{\perp 00}^{ei}$). 
This does not violate energy conservation, as energy conservation ($\nu_{ij}^{ie} = \nu_{ji}^{ei}$) implies 
\begin{equation}
\label{eq:e_cons}
    \frac{1}{3}\nu_{\parallel 00}^{ie} + \frac{2}{3}\nu_{\perp 00}^{ie} = \frac{1}{3}\nu_{\parallel 00}^{ei} + \frac{2}{3}\nu_{\perp 00}^{ei}.
\end{equation} 
However, it means that electron-ion and ion-electron relaxation rates are not inverse processes in each the parallel and perpendicular directions individually, as they are in a weakly magnetized plasma. 
Each of these symmetry breaking effects leads to the formation of a temperature anisotropy during the relaxation process in the strongly magnetized regime ($\beta_e \gtrsim 1$).

The most stark trend in Fig.~\ref{fig:relaxrates}(b) is that the perpendicular relaxation rate decreases rapidly when electrons become strongly magnetized. When the magnetization strength of $\beta_e$ is equal to ten, it becomes two orders of magnitude smaller than the parallel component, indicating that the perpendicular kinetic energy becomes an adiabatic invariant. This transition occurs at a smaller magnetization strength for electrons ($\nu_{\perp 00}^{ei})$ than for ions ($\nu_{\perp 00}^{ie})$ because $\beta_i/\beta_e = \sqrt{m_e/m_i} \ll 1$ (assuming $n_e=n_i$). 
An implication is that at high magnetization strengths, where $|\nu^{ei}_{\perp 00}| \ll |\nu^{ei}_{\parallel 00}|$, the total energy relaxation rate [Eq.~(\ref{eq:e_cons})] is dominated by the parallel component: $\nu^{ei}_{00} \approx \nu^{ei}_{\parallel 00}/3$.

Panel (c) of Fig~\ref{fig:relaxrates} shows the parallel and perpendicular relaxation rates of ions due to collisions with electrons having an anisotropic velocity distribution ($\nu^{ie}_{\parallel 01}$ and $\nu^{ie}_{\perp 01}$). For weak magnetization, the fact that $\nu_{01}^{ie} =0$ and Eq.~(\ref{eq:rate_ptoA}) imply that the parallel and perpendicular relaxation rates follow $\nu^{ie}_{\perp 01}  = -\nu^{ie}_{\parallel 01}/2 $.  
When the plasma is strongly magnetized, the symmetry $\nu^{ie}_{\perp 01}  = -\nu^{ie}_{\parallel 01}/2 $ obeyed in the weakly magnetized regime is broken by the asymmetry in the collision rates.
The parallel relaxation rate becomes positive, increases in magnitude, and reaches a constant value. The perpendicular coefficient first slightly increases in magnitude, then sharply decreases at high magnetization, similar to the trend observed in the case of $\nu^{ie}_{\perp 00}$.
As with $\nu^{ie}_{\perp 00}$, the sharp decrease in $\nu^{ie}_{\perp 01}$ is associated with strong magnetization of the ions ($\beta_i \gtrsim 1$), indicating that the ion perpendicular energy relaxation becomes an adiabatic invariant whether the collisions are with an isotropic, or anisotropic, electron distribution.

Panel (d) of Fig.~\ref{fig:relaxrates} shows the parallel and perpendicular relaxation rates of an anisotropic distribution of electrons colliding with an isotropic distribution of ions ($\nu^{ei}_{\parallel 10}$ and $\nu^{ei}_{\perp 10}$). This term corresponds to the relaxation of electron temperature anisotropy due to collisions with ions. In the weakly magnetized regime, the magnitude of this term is very large compared with the other terms in the electron-ion relaxation rate equation. It is comparable to that of the electron-electron collisional relaxation rate.
In the strongly magnetized regime, both the parallel and perpendicular components decrease rapidly as the electrons become strongly magnetized.  
However, at approximately $\beta_e = 10$, the parallel relaxation rate plateaus to a constant value and $\nu^{ei}_{\perp 10} \approx 0 $. Thus, $\nu^{ei}_{10} =  \nu^{ei}_{\parallel 10}/3$. Energy conservation, $\nu^{ei}_{10} = \nu^{ie}_{01}$, then implies  $\nu^{ei}_{\parallel 10} = 3 \nu^{ie}_{01}$.  At extreme magnetization, ion perpendicular energy becomes an adiabatic invariant and the magnitude of $\nu^{ie}_{\perp 01}$ decreases, while $\nu^{ie}_{\parallel 01}$ remains constant. This reduces the magnitude of $\nu^{ie}_{01}$, which in turn decreases $\nu^{ei}_{\parallel 10}$.


\begin{figure} [!htb] 
\centerline{\includegraphics[width = 8.5cm]{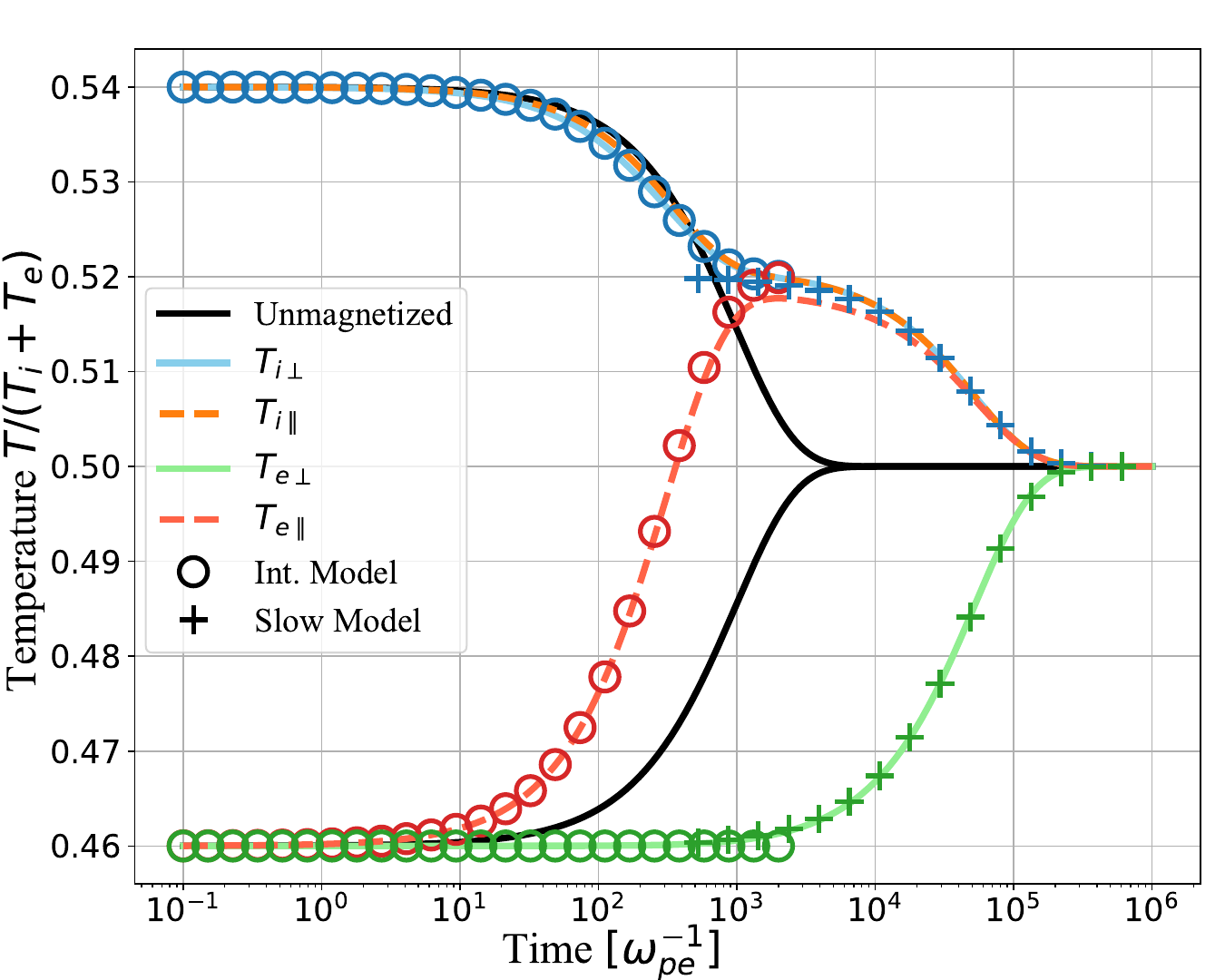}}
\caption {Prediction of the temperature evolution of ions collisionally cooling on electrons from full theoretical model of Sec.~\ref{sec:theory} (colored dashed and solid lines). Here, the coupling strength of the electrons was, $\Gamma_e = 1$, the magnetization strength, $\beta_e = 34$, and densities $n_i=n_e$. For comparison, results for the unmagnetized limit are shown as well (black solid lines). The 'o' markers show the intermediate time scale reduced model from Sec.~\ref{sec:intermediate}. The '+' markers show the long time scale reduced model from Sec.~\ref{sec:slow}. }
  \label{fig:Tevolution}
\end{figure}

\section{Example\label{sec:discu}}

\subsection{Temperature Evolution}

To demonstrate novel aspects of temperature relaxation in strongly magnetized plasmas, we evaluate the model from Sec.~\ref{sec:relaxation rates} for the conditions $\Gamma_e =1$ and $\beta_e=34$. 
This corresponds to a regime where the electrons are strongly magnetized $\beta_e \gg 1$, but the ions are not $\beta_i \approx 1$. 
This regime is relevant to the plasma conditions in experiments on the ALPHA apparatus at CERN ~\cite{Andresen_PRL_2008}. The production rate of trappable antihydrogen in ALPHA improves with colder precursor plasma temperatures. Thus sympathetically cooling experiments, such as antiprotons cooling on electrons~\cite{Andresen_PRL_2008} or positrons cooling on laser-cooled positive Beryllium ions~\cite{Baker_Nature_Comm_2021}, are of interest. These experiments use Penning-Malmberg traps operating in the strongly magnetized regimes $(\beta\geq1)$ and often involve two species with the same charge. The operating parameters for non-neutral plasmas can span a large regime: number densities range from $10^{14}$~m$^{-3}$ to $10^{16}$~m$^{-3}$, temperatures range from 7~K to 100~K, magnetic field strengths typically range from 1~T to 3~T ~\cite{Fajans_POP_2020,Baker_Nature_Comm_2021,Andresen_PRL_2008}. In terms of magnetization strength and coupling strength, $\beta$ ranges from $22$ to $310$ over even higher in some experiments, while $\Gamma$ range from $0.07$ to $0.7$.



Predictions for the temperature evolution of ions and electrons computed from the full temperature evolution model described in Sec.~\ref{sec:theory} are shown in Fig.~\ref{fig:Tevolution}. Both species were initialized with isotropic distributions with the same density and slightly different temperatures. 
The resulting temperature evolution can be described by three characteristic relaxation timescales. 
The fastest timescale is associated with ion-ion collisions, which are well described by the conventional isotropic scattering model associated with the weakly magnetized limit since ions are not strongly magnetized. 
Because this process is so fast, no ion anisotropy forms, as indicated by the overlapping blue and orange dashed lines throughout the evolution.  
The second timescale is associated with relaxation of the parallel electron temperature with the ion temperature, which occurs over approximately $10^3\omega_{pe}^{-1}$. 
The perpendicular electron temperature remains unchanged over this timescale, indicating that it remains an adiabatic invariant. 
The third, and by far the longest, relaxation timescale is the relaxation of the perpendicular electron temperature with the others; lasting over $10^5 \omega_{pe}^{-1}$.
Because it is so slow, the perpendicular electron temperature relaxation process determines the overall temperature relaxation rate. 
The following subsections provide reduced models that focus on each of these regimes. 
 
\subsection{Shortest timescale: Ion anisotropy relaxation}

Since the ions are weakly magnetized at these conditions ($\beta_i \approx 1$), the suppression of ion-ion perpendicular energy relaxation that is expected at strong ion magnetization does not apply. 
Considering the terms of Eq.~(\ref{eq:DTi_eq}) where the coefficients of Eq.~(\ref{eq:weakB_rates}) apply, the ion-ion relaxation process associated with $\nu^{ii}$ is expected to be larger than the others by a factor of $\sqrt{m_i/m_e}$. 
In this case, Eq.~(\ref{eq:DTi_eq}) reduces to
\begin{eqnarray}
    \frac{d \Delta T_i}{d t} = -3 \nu^{ii} \Delta T_i ,
\end{eqnarray}
which has a simple exponential solution 
\begin{equation}
    \Delta T_i (t) = \Delta T_i(0) \exp(-3 \nu^{ii}t) .
\end{equation}
The fact that $\nu^{ii}$ is larger than any other relaxation rate indicates that the fastest process is the relaxation of ion temperature anisotropy. 
In practice, since the ions are initiated with an isotropic distribution function, this simply implies that no ion anisotropy develops at any time during the evolution. 
This greatly simplifies the analysis, since $T_{i\parallel} = T_{i\perp} = T_i$ can be assumed at all times. 
It should be noted, however, that this result is only expected when ions are weakly magnetized. 
If $\beta_i \gg 1$, strong magnetization can suppress the perpendicular energy relaxation in ion-ion collisions dramatically; enough to exceed the mass ratio scaling that usually makes anisotropy relaxation the fastest process. 

\subsection{Intermediate timescale: $T_{e\parallel}$ relaxation\label{sec:intermediate}}

The second fastest process is the equilibration of the electron parallel temperature with the ion temperature. During this stage, the electron perpendicular temperature is constant ($T_{e\perp} = T_{e\perp 0}$) because the collision processes associated with perpendicular electron energy exchange ($\nu_{\perp 00}^{ei}$, $\nu_{\perp 10}^{ei}$ and $\nu^{ee}$) are small; see Fig.~\ref{fig:relaxrates} and Eq.~(\ref{dTe_perpdt}). 
The temperature anisotropy of ions is also negligible  ($\Delta T_{i}=0$). 
Equation~(\ref{eq:rate_ptoA}) then implies that $\nu_{A00}^{ei} = -3\nu_{00}^{ie}$ and $\nu_{A10}^{ie} = -3\nu_{01}^{ie}$. 
Similarly, $\chi_{\perp 10}^{ei}$ is small, so Eq.~(\ref{eq:chi_perp}) implies $\chi_{A10}^{ei} = \chi_{01}^{ie}$. 
Modeling the temperature evolution in this intermediate stage using Eq.~(\ref{eq:rates_1}), these approximations imply that 
\begin{subequations}
    \begin{align}
        \frac{dT_i}{dt} &= -\nu_{00}^{ie}(T_i-T_e) + \nu_{01}^{ie}(1 + \chi_{01}^{ie}) \Delta T_e 
        \label{eq:int_T1} \\
        \frac{dT_e}{dt} &= -\nu_{00}^{ie}(T_e - T_i) -\nu_{01}^{ie}(1+\chi_{01}^{ie})\Delta T_e , 
        \label{eq:int_T2}
    \end{align}
\end{subequations}
while Eq.~(\ref{eq:DTi_eq}) is irrelevant since $\Delta T_i =0$, and Eq.~(\ref{DTe_Eqn}) is redundant with Eq.~(\ref{eq:Te_eq}). 

The sum of Eqs.~(\ref{eq:int_T1}) and (\ref{eq:int_T2}) simply express energy conservation, $T_e + T_i = T_{eo} + T_{io}$. 
At the end of the intermediate stage $T_{e\parallel} = T_i = \tilde{T}$, so energy conservation implies that 
\begin{equation}
    \tilde{T} = \frac{1}{4} T_{eo} + \frac{3}{4} T_{io}. 
\end{equation}
The long-time limit of either Eq.~(\ref{eq:int_T1}) or (\ref{eq:int_T2}) then imply that 
\begin{equation}
\label{eq:long_t}
    \nu_{00}^{ie} = \frac{3}{2} \nu_{01}^{ie}(1 + \chi_{01}^{ie}) .
\end{equation}
With these, Eqs.~(\ref{eq:int_T1}) and (\ref{eq:int_T2}) can be solved analytically showing a simple exponential temperature evolution 
\begin{equation}
\label{eq:Te_par_int}
    T_{e\parallel}(t) = \tilde{T} + \frac{3}{4} (T_{eo} - T_{io}) e^{-4\nu_{00}^{ie} t}
\end{equation}
and 
\begin{equation}
\label{eq:Ti_int}
    T_i(t) = \tilde{T} - \frac{1}{4} (T_{eo} - T_{io}) e^{-4\nu_{00}^{ie} t} .
\end{equation}
Solutions of Eqs.~(\ref{eq:Te_par_int}) and (\ref{eq:Ti_int}) are compared with a full numerical solution of the temperature relaxation problem in Fig.~\ref{fig:Tevolution}. 
This shows excellent agreement over the intermediate timescale regime, which ranges from the initial condition to approximately $10^3\omega_{pe}^{-1}$ at these conditions. 

It is noteworthy that the intermediate relaxation timescale depends only on a single relaxation rate $\nu_{00}^{ie}$, rather than the 11 different coefficients arising in the general temperature evolution. 
This greatly simplifies the relaxation problem. 
It may allow one to isolate and measure this coefficient using an experiment or molecular dynamics simulation in order to test the theoretical prediction. 



\subsection{Slowest timescale: $T_{e\perp}$ relaxation\label{sec:slow}}

The slowest process is the equilibration of the perpendicular electron temperature with the rest of the system. During this stage, the ion temperature and parallel electron temperature are approximately equal $T_i = T_{e\parallel}$ and there is no ion temperature anisotropy $\Delta T_i = 0$. 
The relevant evolution equation in this limit is Eq.~(\ref{dTe_perpdt}), which reduces to 
\begin{equation}
\label{eq:Te_perp_long}
    \frac{d T_{e\perp}}{dt} \approx \nu^{ee} \Delta T_e
\end{equation}
after making use of the above assumptions and the fact that the smallness of $\nu_{\perp ij}^{ei}$ makes these coefficients negligible compared to $\nu^{ee}$. Using energy conservation $(dT_i/dt = -dT_e/dt)$ and $T_i=T_{e\parallel}$, the evolution of the ion and electron parallel temperature reduces to
\begin{equation}
\label{eq:Ti_Te_para_long}
    \frac{d T_{e\parallel}}{dt} = \frac{dT_i}{dt} \approx - \frac{1}{2} \nu^{ee} \Delta T_e .
\end{equation}
Solving Eq.~(\ref{eq:Te_perp_long}) and Eq.~(\ref{eq:Ti_Te_para_long}), applying the initial condition for the long-time stage from $T_{e\parallel} = T_i = \tilde{T}$ and $T_{e\perp} = T_{eo}$,
the solution for the ion and electron temperatures in the long-time limit are
\begin{equation}
\label{eq:Ti_long}
    T_i(t) = \frac{1}{2}(T_{io} + T_{eo}) + \frac{1}{4} (T_{io} - T_{eo}) e^{-\frac{3}{2} \nu^{ee} t} , 
\end{equation}
and 
\begin{equation}
\label{eq:Te_long}
    T_{e\perp}(t) = \frac{1}{2} (T_{io} + T_{eo}) - \frac{1}{2} (T_{io} - T_{eo}) e^{-\frac{3}{2}\nu^{ee} t} .
\end{equation}

Figure~\ref{fig:Tevolution} shows excellent agreement between the analytic solutions from Eqs.~(\ref{eq:Ti_long}) and (\ref{eq:Te_long}) and the numerical solution of the full temperature evolution equation. 
This corresponds to times from approximately $10^3 - 10^5 \omega_{pe}^{-1}$. 
Again, this demonstrates a great simplification of the full evolution process if one focuses on the long-time limit. 
    In this limit, the overall temperature relaxation is dominated by the electron-electron temperature anisotropy relaxation. 
This may enable a future experiment or molecular dynamics simulation to focus on measuring the electron-electron temperature anisotropy rate.

\section{Conclusion}

In this work, we used the recently developed generalized Boltzmann kinetic theory to calculate the temperature relaxation rate of ions and electrons. A time evolution equation was obtained for the temperatures and temperature anisotropies of both the ions and electrons. This was achieved by expanding the distribution function of ions and electrons to the first order in the temperature anisotropy. 

The temperature evolution equations were solved for the case of ions and electrons having different initial temperatures for the conditions $\Gamma_e =1 $ and $\beta_e = 34$. Since the ions are weakly magnetized at these conditions, the temperature anisotropy relaxation of ions due to ion-ion collisions was found to be the fastest process. In the second stage of evolution,  electron parallel temperature and ion temperature come to a common value. The magnetic field is found to suppress the perpendicular energy exchange rate of electrons strongly, and thus, the perpendicular temperature of the electrons is an adiabatic invariant at this stage. In the final stage of the temperature evolution, the electron perpendicular and other temperatures ($T_{i}, T_{e\parallel}$) slowly reach equilibrium at a rate determined by electron perpendicular energy exchange density.

This work focused on repulsive collisions between ions and electrons, which has direct application in the ALPHA experiment, where the antiprotons are collisionally cooled with electrons in a Penning trap~\cite{Andresen_PRL_2008}. However, modeling the temperature evolution in other strongly magnetized systems like ultracold-neutral plasmas and magnetized dusty plasma experiments must account for the attractive collisions~\cite{Guthrie_POP_2021, Pak_PRE_2024, Thomas_PPCF_2012}. This will be explored in future works.

\section{Data Availability Statement}

The data that support the findings of this study are available from the corresponding author upon reasonable request.

\section{Author Declarations}
\subsection{Conflict of Interest}
The authors have no conflicts to disclose.
 
\begin{acknowledgments}
The authors thank Lucas J. Babati and Julian P. Kinney for helpful conversations during the preparation of the manuscript.
This material is based upon work supported by  NSF grant awards No.~PHY-2205506 and PHY-2205620.  
It used Expanse at San Diego Supercomputer Center through allocation PHY-150018 from the Advanced Cyberinfrastructure Coordination Ecosystem: Services \& Support (ACCESS) program, which is supported by National Science Foundation grants \#2138259, \#2138286, \#2138307, \#2137603, and \#2138296.
\end{acknowledgments}

\appendix

\section{Rate coefficient definitions\label{app:rates}}

Definitions for the rate coefficients in Eq.~(\ref{eq:rates_1}) are
\begin{subequations}
\begin{eqnarray}
    \nu^{ie}_{00}&=&\nu^{ei}_{00} = \frac{2 \mathcal{\tilde{P}}^{ie}_{00}}{3 T_e} \\
    \nu^{ie}_{10} &=&\nu^{ei}_{01} = \frac{-2\mathcal{\tilde{P}}^{ie}_{A00}}{9 T_e} \\
    \nu^{ie}_{01} &=& \nu^{ei}_{10}= \frac{2 \mathcal{\tilde{P}}^{ie}_{01} }{3}  \\
    \nu^{ ie}_{A00} &=&  \frac{-\mathcal{\tilde{P}}^{ie}_{A00}}{T_e} \\
    \nu^{ ie}_{A10} &=&   \frac{ \left( 2\mathcal{\tilde{P}}^{ie}_{00}+\mathcal{\tilde{P}}^{ie}_{A00} \right)}{3 T_e} \\
    \nu^{ ie}_{A01} &=&  \frac{\mathcal{-\tilde{P}}^{ie}_{A01}}{T_e} \\
    \nu^{ ei}_{A00} &=&  \frac{\mathcal{\tilde{P}}^{ei}_{A00}}{T_e} \\
    \nu^{ ei}_{A01} &=&  \frac{\mathcal{\tilde{P}}^{ei}_{00}  }{T_e} =-\frac{3}{2} \nu^{ie}_{00}\\
    \nu^{ ei}_{A10} &=&  \frac{\mathcal{\tilde{P}}^{ei}_{A10}}{T_e}. 
\end{eqnarray}
\end{subequations}
Since $\nu^{ie}_{00}$, $\nu^{ie}_{10}$, $\nu^{ie}_{A00}$, $\nu^{ie}_{A10}$ depends only on $\mathcal{\tilde{P}}^{ie}_{00}$ and $\mathcal{\tilde{P}}^{ie}_{A00}$, they are related as
\begin{subequations}
    \begin{eqnarray}
      \nu^{ie}_{10} &=& \frac{2\nu^{ie}_{A00}}{9}  \label{apnuie10}\\
      \nu^{ie}_{A10} &=& \nu^{ie}_{00}-\frac{\nu^{ie}_{A00}}{3} \label{apnuaie10}
    \end{eqnarray}
\end{subequations}

The rate coefficients $\mu$ in terms of $\nu$ and $\chi$.
\begin{subequations}
\begin{eqnarray}
    \mu^{ie}_{00} &=&  \nu^{ie}_{00}\\
    \mu^{ie}_{10} &=&  \nu^{ie}_{10} \\
    \mu^{ie}_{01} &=&  \nu^{ie}_{01} \left( 1+\chi^{ie}_{01}  \frac{T_i}{T_e} \right)\\
    \mu^{ie}_{A00} &=&  \nu^{ie}_{A00}\\
    \mu^{ie}_{A10} &=&  -\nu^{ie}_{A10}\\
    \mu^{ie}_{A01} &=&  -\nu^{ie}_{A01} \left( 1+\chi^{ie}_{A01}  \frac{T_i}{T_e} \right)\\
    \mu^{ei}_{00} &=& \frac{n_i}{n_e} \nu^{ei}_{00}\\
    \mu^{ei}_{10} &=& \frac{n_i}{n_e} \nu^{ei}_{10}  \left( 1+\chi^{ei}_{10}  \frac{T_i}{T_e} \right)\\
    \mu^{ei}_{01} &=& \frac{n_i}{n_e}\nu^{ei}_{01} \\   
    \mu^{ei}_{A00} &=& \frac{n_i}{n_e} \nu^{ei}_{A00}\\
    \mu^{ei}_{A01} &=& \frac{n_i}{n_e} \nu^{ei}_{A01} \chi^{ei}_{A01} \\   
    \mu^{ei}_{A10} &=& \frac{n_i}{n_e} \nu^{ei}_{A10} \left( 1+ \chi^{ei}_{A10}   \frac{T_i}{T_e} \right)\\
    \mu^{ii} &=&  3\nu^{ii} \\
    \mu^{ee} &=&  3\nu^{ee}
\end{eqnarray}
\end{subequations}

\section{Weakly magnetized results \label{sec:weak_results}}
When the plasma is weakly magnetized, the generalized Boltzmann kinetic theory reduces to the traditional Boltzmann kinetic theory~\cite{Jose_POP_2020}. In this limit, the surface integral in the collision operator can be replaced with the differential scattering cross section, which simplifies the integral, and analytical expressions for energy exchange densities can be obtained. The energy exchange densities obtained using the traditional Boltzmann theory are
\begin{widetext}
\begin{subequations}
\begin{eqnarray}
    \mathcal{Q}^{ie}_{\parallel 00}/\tilde{\mathcal{Q}} &=& \frac{T_e-T_i}{3}, \\
    \mathcal{Q}^{ie}_{\perp 00}/\tilde{\mathcal{Q}} &=& \frac{2 (T_e-T_i)}{3} ,\\
    \mathcal{Q}^{ie}_{\parallel 10}/\tilde{\mathcal{Q}} &=& -\frac{ (10 \Xi^{(1,1)} m_i T_e (m_e+m_i)+4 \Xi^{(1,2)} m_e m_i (T_i-T_e)+3 \Xi^{(2,2)} m_e (m_e T_i+m_i T_e))T_i}{45 \Xi^{(1,1)} m_i (m_e T_i+m_i T_e)} ,\\
    \mathcal{Q}^{ie}_{\perp 10}/\tilde{\mathcal{Q}} &=& \frac{ (10 \Xi^{(1,1)} m_i T_e (m_e+m_i)+4 \Xi^{(1,2)} m_e m_i (T_i-T_e)+3 \Xi^{(2,2)} m_e (m_e T_i+m_i T_e)) T_i}{45 \Xi^{(1,1)} m_i (m_e T_i+m_i T_e)} ,\\
    \mathcal{Q}^{ie}_{\parallel 01}/\tilde{\mathcal{Q}} &=& \frac{ (10 \Xi^{(1,1)} T_i (m_e+m_i)+4 \Xi^{(1,2)} m_i (T_e-T_i)-3 \Xi^{(2,2)} (m_e T_i+m_i T_e)) T_e}{45 \Xi^{(1,1)} (m_e T_i+m_i T_e)},\\
    \mathcal{Q}^{ie}_{\perp 01}/\tilde{\mathcal{Q}} &=& \frac{ (-10 \Xi^{(1,1)} T_i (m_e+m_i)+4 \Xi^{(1,2)} m_i (T_i-T_e)+3 \Xi^{(2,2)} (m_e T_i+m_i T_e)) T_e }{45 \Xi^{(1,1)} (m_e T_i+m_i T_e)},
    \end{eqnarray}
\end{subequations}
where $
  \tilde{\mathcal{Q}} = 4 \sqrt{2 \pi } e^4 \Xi^{(1,1)} n_e n_i \sqrt{m_e m_i}/(m_e T_i+m_i T_e)^{3/2}
$ is a reference energy exchange density. The electron-ion energy exchange densities are
\begin{subequations}
\begin{eqnarray}
    \mathcal{Q}^{ei}_{\parallel 00}/\tilde{\mathcal{Q}} &=& -\frac{T_e-T_i}{3}, \\
    \mathcal{Q}^{ei}_{\perp 00}/\tilde{\mathcal{Q}} &=&  -\frac{2 (T_e-T_i)}{3},   \\
    \mathcal{Q}^{ei}_{\parallel 10}/\tilde{\mathcal{Q}} &=& -\frac{ (10 \Xi^{(1,1)} m_e T_i (m_e+m_i)+4 \Xi^{(1,2)} m_e m_i (T_e-T_i)+3 \Xi^{(2,2)} m_i (m_e T_i+m_i T_e)) T_e}{45 \Xi^{(1,1)} m_e (m_e T_i+m_i T_e)},\\
    \mathcal{Q}^{ei}_{\perp 10}/\tilde{\mathcal{Q}} &=& \frac{ (10 \Xi^{(1,1)} m_e T_i (m_e+m_i)+4 \Xi^{(1,2)} m_e m_i (T_e-T_i)+3 \Xi^{(2,2)} m_i (m_e T_i+m_i T_e))T_e}{45 \Xi^{(1,1)} m_e (m_e T_i+m_i T_e)},\\
    \mathcal{Q}^{ei}_{\parallel 01}/\tilde{\mathcal{Q}} &=& \frac{  (10 \Xi^{(1,1)} T_e (m_e+m_i)+4 \Xi^{(1,2)} m_e (T_i-T_e)-3 \Xi^{(2,2)} (m_e T_i+m_i T_e))T_i}{45 \Xi^{(1,1)} (m_e T_i+m_i T_e)},\\
    \mathcal{Q}^{ei}_{\perp 01}/\tilde{\mathcal{Q}} &=& \frac{ (-10 \Xi^{(1,1)} T_e (m_e+m_i)+4 \Xi^{(1,2)} m_e (T_e-T_i)+3 \Xi^{(2,2)} (m_e T_i+m_i T_e))T_i}{45 \Xi^{(1,1)} (m_e T_i+m_i T_e)} .
\end{eqnarray}
\end{subequations}
The energy exchange densities obey the following relations
\begin{subequations}
\begin{eqnarray}
    \mathcal{Q}^{ie}_{\perp 00} &=& 2\mathcal{Q}^{ie}_{\parallel 00},\\
    \mathcal{Q}^{ie}_{\perp 01} &=& -\mathcal{Q}^{ie}_{\parallel 01},\\
    \mathcal{Q}^{ie}_{\perp 10} &=& -\mathcal{Q}^{ie}_{\parallel 10},\\
    \mathcal{Q}^{ei}_{\perp 00} &=& 2\mathcal{Q}^{ei}_{\parallel 00},\\
    \mathcal{Q}^{ei}_{\perp 01} &=& -\mathcal{Q}^{ei}_{\parallel 01},\\
    \mathcal{Q}^{ei}_{\perp 10} &=& -\mathcal{Q}^{ei}_{\parallel 10}. 
\end{eqnarray} 
\end{subequations}
The energy exchange densities obtained can be simplified by focusing on the limiting case that ions are massive compared to electrons. After expanding the energy exchange densities around this limit, following simplified expressions for ion-electron energy exchange densities can be obtained
\begin{subequations}
\begin{eqnarray}
    \mathcal{Q}^{ie}_{\parallel 00}/\tilde{\mathcal{Q}} &=&  \frac{T_e-T_i}{3}+\mathcal{O}\left(\left(\frac{1}{m_r}\right)^1\right),\\
    \mathcal{Q}^{ie}_{\perp 00}/\tilde{\mathcal{Q}} &=& \frac{2 (T_e-T_i)}{3}+O\left(\left(\frac{1}{m_r}\right)^1\right), \\
    \mathcal{Q}^{ie}_{\parallel 10}/\tilde{\mathcal{Q}} &=& -\frac{2  T_i}{9}+\mathcal{O}\left(\left(\frac{1}{m_r}\right)^1\right), \\
    \mathcal{Q}^{ie}_{\perp 10}/\tilde{\mathcal{Q}} &=& \frac{2  T_i}{9}+\mathcal{O}\left(\left(\frac{1}{m_r}\right)^1\right), \\
    \mathcal{Q}^{ie}_{\parallel 01}/\tilde{\mathcal{Q}} &=& \frac{ 10 \Xi^{(1,1)} T_i+4 \Xi^{(1,2)} (T_e-T_i)-3 \Xi^{(2,2)} T_e}{45 \Xi^{(1,1)} }+\mathcal{O}\left(\left(\frac{1}{m_r}\right)^1\right), \\
    \mathcal{Q}^{ie}_{\perp 01}/\tilde{\mathcal{Q}} &=& \frac{ -10 \Xi^{(1,1)} T_i+4 \Xi^{(1,2)} (T_i-T_e)+3 \Xi^{(2,2)} T_e}{45 \Xi^{(1,1)}}+\mathcal{O}\left(\left(\frac{1}{m_r}\right)^1\right). 
\end{eqnarray}
\end{subequations}
Here, the high mass ratio limit reference energy exchange density is used, $
   \tilde{\mathcal{Q}} =  4 \sqrt{2 \pi } e^4 \Xi^{(1,1)} m_e n_e n_i/(m_r (m_e T_e)^{3/2}). $
The electron-ion energy exchange densities in this limit are
\begin{subequations}
\begin{eqnarray}
    \mathcal{Q}^{ei}_{\parallel 00}/\tilde{\mathcal{Q}} &=&  \frac{T_i-T_e}{3}+\mathcal{O}\left(\left(\frac{1}{m_r}\right)^1\right),\\
    \mathcal{Q}^{ei}_{\perp 00}/\tilde{\mathcal{Q}} &=&-\frac{2 (T_e-T_i)}{3}+\mathcal{O}\left(\left(\frac{1}{m_r}\right)^1\right),\\
    \mathcal{Q}^{ei}_{\parallel 10}/\tilde{\mathcal{Q}} &=& -\frac{ T_e m_r \Xi^{(2,2)} }{15 \Xi^{(1,1)}}+\frac{  -20 \Xi^{(1,1)} T_i+8 \Xi^{(1,2)} (T_i-T_e)+9 \Xi^{(2,2)} T_i}{90 \Xi^{(1,1)} }+\mathcal{O}\left(\sqrt{\frac{1}{m_r}}\right), \\
    \mathcal{Q}^{ei}_{\perp 10}/\tilde{\mathcal{Q}} &=& \frac{ T_e m_r \Xi^{(2,2)} }{15 \Xi^{(1,1)}}+\frac{  20 \Xi^{(1,1)} T_i+8 \Xi^{(1,2)} (T_e-T_i)-9 \Xi^{(2,2)} T_i}{90 \Xi^{(1,1)} }+\mathcal{O}\left(\sqrt{\frac{1}{m_r}}\right), \\
    \mathcal{Q}^{ei}_{\parallel 01}/\tilde{\mathcal{Q}} &=& \frac{T_i}{45}  \left(10-\frac{3 \Xi^{(2,2)}}{\Xi^{(1,1)}}\right)+\mathcal{O}\left(\left(\frac{1}{m_r}\right)^1\right), \\
    \mathcal{Q}^{ei}_{\perp 01}/\tilde{\mathcal{Q}} &=& \frac{T_i}{45}   \left(\frac{3 \Xi^{(2,2)}}{\Xi^{(1,1)}}-10\right)+\mathcal{O}\left(\left(\frac{1}{m_r}\right)^1\right).
\end{eqnarray}
\end{subequations}
The traditional results for weakly coupled plasmas  in terms of Coulomb logarithm ($\ln \Lambda$) can be obtained from the generalized Coulomb logarithm ($\Xi^{(l,k)}$) using the following relations~\cite{Baalrud_POP_2014}: $\Xi^{(1,1)} = \Xi^{(1,2)} = \ln \Lambda$,  and $\Xi^{(2,2)} = 2 \ln \Lambda$.

\end{widetext}

\bibliography{references}	

\begin{thebibliography}{53}%
\makeatletter
\providecommand \@ifxundefined [1]{%
 \@ifx{#1\undefined}
}%
\providecommand \@ifnum [1]{%
 \ifnum #1\expandafter \@firstoftwo
 \else \expandafter \@secondoftwo
 \fi
}%
\providecommand \@ifx [1]{%
 \ifx #1\expandafter \@firstoftwo
 \else \expandafter \@secondoftwo
 \fi
}%
\providecommand \natexlab [1]{#1}%
\providecommand \enquote  [1]{``#1''}%
\providecommand \bibnamefont  [1]{#1}%
\providecommand \bibfnamefont [1]{#1}%
\providecommand \citenamefont [1]{#1}%
\providecommand \href@noop [0]{\@secondoftwo}%
\providecommand \href [0]{\begingroup \@sanitize@url \@href}%
\providecommand \@href[1]{\@@startlink{#1}\@@href}%
\providecommand \@@href[1]{\endgroup#1\@@endlink}%
\providecommand \@sanitize@url [0]{\catcode `\\12\catcode `\$12\catcode `\&12\catcode `\#12\catcode `\^12\catcode `\_12\catcode `\%12\relax}%
\providecommand \@@startlink[1]{}%
\providecommand \@@endlink[0]{}%
\providecommand \url  [0]{\begingroup\@sanitize@url \@url }%
\providecommand \@url [1]{\endgroup\@href {#1}{\urlprefix }}%
\providecommand \urlprefix  [0]{URL }%
\providecommand \Eprint [0]{\href }%
\providecommand \doibase [0]{http://dx.doi.org/}%
\providecommand \selectlanguage [0]{\@gobble}%
\providecommand \bibinfo  [0]{\@secondoftwo}%
\providecommand \bibfield  [0]{\@secondoftwo}%
\providecommand \translation [1]{[#1]}%
\providecommand \BibitemOpen [0]{}%
\providecommand \bibitemStop [0]{}%
\providecommand \bibitemNoStop [0]{.\EOS\space}%
\providecommand \EOS [0]{\spacefactor3000\relax}%
\providecommand \BibitemShut  [1]{\csname bibitem#1\endcsname}%
\let\auto@bib@innerbib\@empty
\bibitem [{\citenamefont {Ott}\ \emph {et~al.}(2017)\citenamefont {Ott}, \citenamefont {Bonitz}, \citenamefont {Hartmann},\ and\ \citenamefont {Donk\'o}}]{Ott_PRE_2017}%
  \BibitemOpen
  \bibfield  {author} {\bibinfo {author} {\bibfnamefont {T.}~\bibnamefont {Ott}}, \bibinfo {author} {\bibfnamefont {M.}~\bibnamefont {Bonitz}}, \bibinfo {author} {\bibfnamefont {P.}~\bibnamefont {Hartmann}}, \ and\ \bibinfo {author} {\bibfnamefont {Z.}~\bibnamefont {Donk\'o}},\ }\href {\doibase 10.1103/PhysRevE.95.013209} {\bibfield  {journal} {\bibinfo  {journal} {Phys. Rev. E}\ }\textbf {\bibinfo {volume} {95}},\ \bibinfo {pages} {013209} (\bibinfo {year} {2017})}\BibitemShut {NoStop}%
\bibitem [{\citenamefont {Baalrud}\ and\ \citenamefont {Daligault}(2017)}]{Baalrud_PRE_2017}%
  \BibitemOpen
  \bibfield  {author} {\bibinfo {author} {\bibfnamefont {S.~D.}\ \bibnamefont {Baalrud}}\ and\ \bibinfo {author} {\bibfnamefont {J.}~\bibnamefont {Daligault}},\ }\href {\doibase 10.1103/PhysRevE.96.043202} {\bibfield  {journal} {\bibinfo  {journal} {Phys. Rev. E}\ }\textbf {\bibinfo {volume} {96}},\ \bibinfo {pages} {043202} (\bibinfo {year} {2017})}\BibitemShut {NoStop}%
\bibitem [{\citenamefont {Ott}\ and\ \citenamefont {Bonitz}(2011)}]{ott2011diffusion}%
  \BibitemOpen
  \bibfield  {author} {\bibinfo {author} {\bibfnamefont {T.}~\bibnamefont {Ott}}\ and\ \bibinfo {author} {\bibfnamefont {M.}~\bibnamefont {Bonitz}},\ }\href@noop {} {\bibfield  {journal} {\bibinfo  {journal} {Physical review letters}\ }\textbf {\bibinfo {volume} {107}},\ \bibinfo {pages} {135003} (\bibinfo {year} {2011})}\BibitemShut {NoStop}%
\bibitem [{\citenamefont {K\"ahlert}\ and\ \citenamefont {Bonitz}(2022)}]{Kahlert_Phys_Rev_Res_2022}%
  \BibitemOpen
  \bibfield  {author} {\bibinfo {author} {\bibfnamefont {H.}~\bibnamefont {K\"ahlert}}\ and\ \bibinfo {author} {\bibfnamefont {M.}~\bibnamefont {Bonitz}},\ }\href {\doibase 10.1103/PhysRevResearch.4.013197} {\bibfield  {journal} {\bibinfo  {journal} {Phys. Rev. Res.}\ }\textbf {\bibinfo {volume} {4}},\ \bibinfo {pages} {013197} (\bibinfo {year} {2022})}\BibitemShut {NoStop}%
\bibitem [{\citenamefont {Kählert}\ and\ \citenamefont {Bonitz}(2023)}]{Kahlert_CPP_2023}%
  \BibitemOpen
  \bibfield  {author} {\bibinfo {author} {\bibfnamefont {H.}~\bibnamefont {Kählert}}\ and\ \bibinfo {author} {\bibfnamefont {M.}~\bibnamefont {Bonitz}},\ }\href@noop {} {\bibfield  {journal} {\bibinfo  {journal} {Contributions to Plasma Physics}\ }\textbf {\bibinfo {volume} {63}},\ \bibinfo {pages} {e202200185} (\bibinfo {year} {2023})}\BibitemShut {NoStop}%
\bibitem [{\citenamefont {Fajans}\ and\ \citenamefont {Surko}(2020)}]{Fajans_POP_2020}%
  \BibitemOpen
  \bibfield  {author} {\bibinfo {author} {\bibfnamefont {J.}~\bibnamefont {Fajans}}\ and\ \bibinfo {author} {\bibfnamefont {C.}~\bibnamefont {Surko}},\ }\href@noop {} {\bibfield  {journal} {\bibinfo  {journal} {Physics of Plasmas}\ }\textbf {\bibinfo {volume} {27}},\ \bibinfo {pages} {030601} (\bibinfo {year} {2020})}\BibitemShut {NoStop}%
\bibitem [{\citenamefont {Stenson}\ \emph {et~al.}(2017)\citenamefont {Stenson}, \citenamefont {Horn-Stanja}, \citenamefont {Stoneking},\ and\ \citenamefont {Pedersen}}]{Stenson_JPP_2017}%
  \BibitemOpen
  \bibfield  {author} {\bibinfo {author} {\bibfnamefont {E.~V.}\ \bibnamefont {Stenson}}, \bibinfo {author} {\bibfnamefont {J.}~\bibnamefont {Horn-Stanja}}, \bibinfo {author} {\bibfnamefont {M.~R.}\ \bibnamefont {Stoneking}}, \ and\ \bibinfo {author} {\bibfnamefont {T.~S.}\ \bibnamefont {Pedersen}},\ }\href {\doibase 10.1017/S0022377817000022} {\bibfield  {journal} {\bibinfo  {journal} {Journal of Plasma Physics}\ }\textbf {\bibinfo {volume} {83}},\ \bibinfo {pages} {595830106} (\bibinfo {year} {2017})}\BibitemShut {NoStop}%
\bibitem [{\citenamefont {Ahmadi}\ \emph {et~al.}(2017)\citenamefont {Ahmadi}, \citenamefont {Alves}, \citenamefont {Baker}, \citenamefont {Bertsche}, \citenamefont {Butler}, \citenamefont {Capra}, \citenamefont {Carruth}, \citenamefont {Cesar}, \citenamefont {Charlton}, \citenamefont {Cohen}, \citenamefont {Collister}, \citenamefont {Eriksson}, \citenamefont {Evans}, \citenamefont {Evetts}, \citenamefont {Fajans}, \citenamefont {Friesen}, \citenamefont {Fujiwara}, \citenamefont {Gill}, \citenamefont {Gutierrez}, \citenamefont {Hangst}, \citenamefont {Hardy}, \citenamefont {Hayden}, \citenamefont {Isaac}, \citenamefont {Ishida}, \citenamefont {Johnson}, \citenamefont {Jones}, \citenamefont {Jonsell}, \citenamefont {Kurchaninov}, \citenamefont {Madsen}, \citenamefont {Mathers}, \citenamefont {Maxwell}, \citenamefont {McKenna}, \citenamefont {Menary}, \citenamefont {Michan}, \citenamefont {Momose}, \citenamefont {Munich}, \citenamefont {Nolan}, \citenamefont {Olchanski}, \citenamefont {Olin}, \citenamefont
  {Pusa}, \citenamefont {Rasmussen}, \citenamefont {Robicheaux}, \citenamefont {Sacramento}, \citenamefont {Sameed}, \citenamefont {Sarid}, \citenamefont {Silveira}, \citenamefont {Stracka}, \citenamefont {Stutter}, \citenamefont {So}, \citenamefont {Tharp}, \citenamefont {Thompson}, \citenamefont {Thompson}, \citenamefont {van~der Werf},\ and\ \citenamefont {Wurtele}}]{Ahmadi_nature_comm_2017}%
  \BibitemOpen
  \bibfield  {author} {\bibinfo {author} {\bibfnamefont {M.}~\bibnamefont {Ahmadi}}, \bibinfo {author} {\bibfnamefont {B.~X.~R.}\ \bibnamefont {Alves}}, \bibinfo {author} {\bibfnamefont {C.~J.}\ \bibnamefont {Baker}}, \bibinfo {author} {\bibfnamefont {W.}~\bibnamefont {Bertsche}}, \bibinfo {author} {\bibfnamefont {E.}~\bibnamefont {Butler}}, \bibinfo {author} {\bibfnamefont {A.}~\bibnamefont {Capra}}, \bibinfo {author} {\bibfnamefont {C.}~\bibnamefont {Carruth}}, \bibinfo {author} {\bibfnamefont {C.~L.}\ \bibnamefont {Cesar}}, \bibinfo {author} {\bibfnamefont {M.}~\bibnamefont {Charlton}}, \bibinfo {author} {\bibfnamefont {S.}~\bibnamefont {Cohen}}, \bibinfo {author} {\bibfnamefont {R.}~\bibnamefont {Collister}}, \bibinfo {author} {\bibfnamefont {S.}~\bibnamefont {Eriksson}}, \bibinfo {author} {\bibfnamefont {A.}~\bibnamefont {Evans}}, \bibinfo {author} {\bibfnamefont {N.}~\bibnamefont {Evetts}}, \bibinfo {author} {\bibfnamefont {J.}~\bibnamefont {Fajans}}, \bibinfo {author} {\bibfnamefont {T.}~\bibnamefont
  {Friesen}}, \bibinfo {author} {\bibfnamefont {M.~C.}\ \bibnamefont {Fujiwara}}, \bibinfo {author} {\bibfnamefont {D.~R.}\ \bibnamefont {Gill}}, \bibinfo {author} {\bibfnamefont {A.}~\bibnamefont {Gutierrez}}, \bibinfo {author} {\bibfnamefont {J.~S.}\ \bibnamefont {Hangst}}, \bibinfo {author} {\bibfnamefont {W.~N.}\ \bibnamefont {Hardy}}, \bibinfo {author} {\bibfnamefont {M.~E.}\ \bibnamefont {Hayden}}, \bibinfo {author} {\bibfnamefont {C.~A.}\ \bibnamefont {Isaac}}, \bibinfo {author} {\bibfnamefont {A.}~\bibnamefont {Ishida}}, \bibinfo {author} {\bibfnamefont {M.~A.}\ \bibnamefont {Johnson}}, \bibinfo {author} {\bibfnamefont {S.~A.}\ \bibnamefont {Jones}}, \bibinfo {author} {\bibfnamefont {S.}~\bibnamefont {Jonsell}}, \bibinfo {author} {\bibfnamefont {L.}~\bibnamefont {Kurchaninov}}, \bibinfo {author} {\bibfnamefont {N.}~\bibnamefont {Madsen}}, \bibinfo {author} {\bibfnamefont {M.}~\bibnamefont {Mathers}}, \bibinfo {author} {\bibfnamefont {D.}~\bibnamefont {Maxwell}}, \bibinfo {author} {\bibfnamefont
  {J.~T.~K.}\ \bibnamefont {McKenna}}, \bibinfo {author} {\bibfnamefont {S.}~\bibnamefont {Menary}}, \bibinfo {author} {\bibfnamefont {J.~M.}\ \bibnamefont {Michan}}, \bibinfo {author} {\bibfnamefont {T.}~\bibnamefont {Momose}}, \bibinfo {author} {\bibfnamefont {J.~J.}\ \bibnamefont {Munich}}, \bibinfo {author} {\bibfnamefont {P.}~\bibnamefont {Nolan}}, \bibinfo {author} {\bibfnamefont {K.}~\bibnamefont {Olchanski}}, \bibinfo {author} {\bibfnamefont {A.}~\bibnamefont {Olin}}, \bibinfo {author} {\bibfnamefont {P.}~\bibnamefont {Pusa}}, \bibinfo {author} {\bibfnamefont {C.~{\O}.}\ \bibnamefont {Rasmussen}}, \bibinfo {author} {\bibfnamefont {F.}~\bibnamefont {Robicheaux}}, \bibinfo {author} {\bibfnamefont {R.~L.}\ \bibnamefont {Sacramento}}, \bibinfo {author} {\bibfnamefont {M.}~\bibnamefont {Sameed}}, \bibinfo {author} {\bibfnamefont {E.}~\bibnamefont {Sarid}}, \bibinfo {author} {\bibfnamefont {D.~M.}\ \bibnamefont {Silveira}}, \bibinfo {author} {\bibfnamefont {S.}~\bibnamefont {Stracka}}, \bibinfo {author}
  {\bibfnamefont {G.}~\bibnamefont {Stutter}}, \bibinfo {author} {\bibfnamefont {C.}~\bibnamefont {So}}, \bibinfo {author} {\bibfnamefont {T.~D.}\ \bibnamefont {Tharp}}, \bibinfo {author} {\bibfnamefont {J.~E.}\ \bibnamefont {Thompson}}, \bibinfo {author} {\bibfnamefont {R.~I.}\ \bibnamefont {Thompson}}, \bibinfo {author} {\bibfnamefont {D.~P.}\ \bibnamefont {van~der Werf}}, \ and\ \bibinfo {author} {\bibfnamefont {J.~S.}\ \bibnamefont {Wurtele}},\ }\href {\doibase 10.1038/s41467-017-00760-9} {\bibfield  {journal} {\bibinfo  {journal} {Nature Communications}\ }\textbf {\bibinfo {volume} {8}},\ \bibinfo {pages} {681} (\bibinfo {year} {2017})}\BibitemShut {NoStop}%
\bibitem [{\citenamefont {Baker}\ \emph {et~al.}(2021)\citenamefont {Baker}, \citenamefont {Bertsche}, \citenamefont {Capra}, \citenamefont {Cesar}, \citenamefont {Charlton}, \citenamefont {Mathad}, \citenamefont {Eriksson}, \citenamefont {Evans}, \citenamefont {Evetts}, \citenamefont {Fabbri}, \citenamefont {Fajans}, \citenamefont {Friesen}, \citenamefont {Fujiwara}, \citenamefont {Grandemange}, \citenamefont {Granum}, \citenamefont {Hangst}, \citenamefont {Hayden}, \citenamefont {Hodgkinson}, \citenamefont {Isaac}, \citenamefont {Johnson}, \citenamefont {Jones}, \citenamefont {Jones}, \citenamefont {Jonsell}, \citenamefont {Kurchaninov}, \citenamefont {Madsen}, \citenamefont {Maxwell}, \citenamefont {McKenna}, \citenamefont {Menary}, \citenamefont {Momose}, \citenamefont {Mullan}, \citenamefont {Olchanski}, \citenamefont {Olin}, \citenamefont {Peszka}, \citenamefont {Powell}, \citenamefont {Pusa}, \citenamefont {Rasmussen}, \citenamefont {Robicheaux}, \citenamefont {Sacramento}, \citenamefont {Sameed},
  \citenamefont {Sarid}, \citenamefont {Silveira}, \citenamefont {Stutter}, \citenamefont {So}, \citenamefont {Tharp}, \citenamefont {Thompson}, \citenamefont {van~der Werf},\ and\ \citenamefont {Wurtele}}]{Baker_Nature_Comm_2021}%
  \BibitemOpen
  \bibfield  {author} {\bibinfo {author} {\bibfnamefont {C.~J.}\ \bibnamefont {Baker}}, \bibinfo {author} {\bibfnamefont {W.}~\bibnamefont {Bertsche}}, \bibinfo {author} {\bibfnamefont {A.}~\bibnamefont {Capra}}, \bibinfo {author} {\bibfnamefont {C.~L.}\ \bibnamefont {Cesar}}, \bibinfo {author} {\bibfnamefont {M.}~\bibnamefont {Charlton}}, \bibinfo {author} {\bibfnamefont {A.~C.}\ \bibnamefont {Mathad}}, \bibinfo {author} {\bibfnamefont {S.}~\bibnamefont {Eriksson}}, \bibinfo {author} {\bibfnamefont {A.}~\bibnamefont {Evans}}, \bibinfo {author} {\bibfnamefont {N.}~\bibnamefont {Evetts}}, \bibinfo {author} {\bibfnamefont {S.}~\bibnamefont {Fabbri}}, \bibinfo {author} {\bibfnamefont {J.}~\bibnamefont {Fajans}}, \bibinfo {author} {\bibfnamefont {T.}~\bibnamefont {Friesen}}, \bibinfo {author} {\bibfnamefont {M.~C.}\ \bibnamefont {Fujiwara}}, \bibinfo {author} {\bibfnamefont {P.}~\bibnamefont {Grandemange}}, \bibinfo {author} {\bibfnamefont {P.}~\bibnamefont {Granum}}, \bibinfo {author} {\bibfnamefont {J.~S.}\
  \bibnamefont {Hangst}}, \bibinfo {author} {\bibfnamefont {M.~E.}\ \bibnamefont {Hayden}}, \bibinfo {author} {\bibfnamefont {D.}~\bibnamefont {Hodgkinson}}, \bibinfo {author} {\bibfnamefont {C.~A.}\ \bibnamefont {Isaac}}, \bibinfo {author} {\bibfnamefont {M.~A.}\ \bibnamefont {Johnson}}, \bibinfo {author} {\bibfnamefont {J.~M.}\ \bibnamefont {Jones}}, \bibinfo {author} {\bibfnamefont {S.~A.}\ \bibnamefont {Jones}}, \bibinfo {author} {\bibfnamefont {S.}~\bibnamefont {Jonsell}}, \bibinfo {author} {\bibfnamefont {L.}~\bibnamefont {Kurchaninov}}, \bibinfo {author} {\bibfnamefont {N.}~\bibnamefont {Madsen}}, \bibinfo {author} {\bibfnamefont {D.}~\bibnamefont {Maxwell}}, \bibinfo {author} {\bibfnamefont {J.~T.~K.}\ \bibnamefont {McKenna}}, \bibinfo {author} {\bibfnamefont {S.}~\bibnamefont {Menary}}, \bibinfo {author} {\bibfnamefont {T.}~\bibnamefont {Momose}}, \bibinfo {author} {\bibfnamefont {P.}~\bibnamefont {Mullan}}, \bibinfo {author} {\bibfnamefont {K.}~\bibnamefont {Olchanski}}, \bibinfo {author}
  {\bibfnamefont {A.}~\bibnamefont {Olin}}, \bibinfo {author} {\bibfnamefont {J.}~\bibnamefont {Peszka}}, \bibinfo {author} {\bibfnamefont {A.}~\bibnamefont {Powell}}, \bibinfo {author} {\bibfnamefont {P.}~\bibnamefont {Pusa}}, \bibinfo {author} {\bibfnamefont {C.~{\O}.}\ \bibnamefont {Rasmussen}}, \bibinfo {author} {\bibfnamefont {F.}~\bibnamefont {Robicheaux}}, \bibinfo {author} {\bibfnamefont {R.~L.}\ \bibnamefont {Sacramento}}, \bibinfo {author} {\bibfnamefont {M.}~\bibnamefont {Sameed}}, \bibinfo {author} {\bibfnamefont {E.}~\bibnamefont {Sarid}}, \bibinfo {author} {\bibfnamefont {D.~M.}\ \bibnamefont {Silveira}}, \bibinfo {author} {\bibfnamefont {G.}~\bibnamefont {Stutter}}, \bibinfo {author} {\bibfnamefont {C.}~\bibnamefont {So}}, \bibinfo {author} {\bibfnamefont {T.~D.}\ \bibnamefont {Tharp}}, \bibinfo {author} {\bibfnamefont {R.~I.}\ \bibnamefont {Thompson}}, \bibinfo {author} {\bibfnamefont {D.~P.}\ \bibnamefont {van~der Werf}}, \ and\ \bibinfo {author} {\bibfnamefont {J.~S.}\ \bibnamefont
  {Wurtele}},\ }\href {\doibase 10.1038/s41467-021-26086-1} {\bibfield  {journal} {\bibinfo  {journal} {Nature Communications}\ }\textbf {\bibinfo {volume} {12}},\ \bibinfo {pages} {6139} (\bibinfo {year} {2021})}\BibitemShut {NoStop}%
\bibitem [{\citenamefont {Thomas}\ \emph {et~al.}(2012)\citenamefont {Thomas}, \citenamefont {Merlino},\ and\ \citenamefont {Rosenberg}}]{Thomas_PPCF_2012}%
  \BibitemOpen
  \bibfield  {author} {\bibinfo {author} {\bibfnamefont {E.}~\bibnamefont {Thomas}}, \bibinfo {author} {\bibfnamefont {R.~L.}\ \bibnamefont {Merlino}}, \ and\ \bibinfo {author} {\bibfnamefont {M.}~\bibnamefont {Rosenberg}},\ }\href {\doibase 10.1088/0741-3335/54/12/124034} {\bibfield  {journal} {\bibinfo  {journal} {Plasma Physics and Controlled Fusion}\ }\textbf {\bibinfo {volume} {54}},\ \bibinfo {pages} {124034} (\bibinfo {year} {2012})}\BibitemShut {NoStop}%
\bibitem [{\citenamefont {Beck}\ \emph {et~al.}(1992)\citenamefont {Beck}, \citenamefont {Fajans},\ and\ \citenamefont {Malmberg}}]{Beck_PRL_1992}%
  \BibitemOpen
  \bibfield  {author} {\bibinfo {author} {\bibfnamefont {B.~R.}\ \bibnamefont {Beck}}, \bibinfo {author} {\bibfnamefont {J.}~\bibnamefont {Fajans}}, \ and\ \bibinfo {author} {\bibfnamefont {J.~H.}\ \bibnamefont {Malmberg}},\ }\href {\doibase 10.1103/PhysRevLett.68.317} {\bibfield  {journal} {\bibinfo  {journal} {Phys. Rev. Lett.}\ }\textbf {\bibinfo {volume} {68}},\ \bibinfo {pages} {317} (\bibinfo {year} {1992})}\BibitemShut {NoStop}%
\bibitem [{\citenamefont {Glinsky}\ \emph {et~al.}(1992)\citenamefont {Glinsky}, \citenamefont {O’Neil}, \citenamefont {Rosenbluth}, \citenamefont {Tsuruta},\ and\ \citenamefont {Ichimaru}}]{Glinsky_Phys_Fluids_1992}%
  \BibitemOpen
  \bibfield  {author} {\bibinfo {author} {\bibfnamefont {M.~E.}\ \bibnamefont {Glinsky}}, \bibinfo {author} {\bibfnamefont {T.~M.}\ \bibnamefont {O’Neil}}, \bibinfo {author} {\bibfnamefont {M.~N.}\ \bibnamefont {Rosenbluth}}, \bibinfo {author} {\bibfnamefont {K.}~\bibnamefont {Tsuruta}}, \ and\ \bibinfo {author} {\bibfnamefont {S.}~\bibnamefont {Ichimaru}},\ }\href@noop {} {\bibfield  {journal} {\bibinfo  {journal} {Physics of Fluids B: Plasma Physics}\ }\textbf {\bibinfo {volume} {4}},\ \bibinfo {pages} {1156} (\bibinfo {year} {1992})}\BibitemShut {NoStop}%
\bibitem [{\citenamefont {Gorman}\ \emph {et~al.}(2022)\citenamefont {Gorman}, \citenamefont {Warrens}, \citenamefont {Bradshaw},\ and\ \citenamefont {Killian}}]{Gorman_PRA_2022}%
  \BibitemOpen
  \bibfield  {author} {\bibinfo {author} {\bibfnamefont {G.~M.}\ \bibnamefont {Gorman}}, \bibinfo {author} {\bibfnamefont {M.~K.}\ \bibnamefont {Warrens}}, \bibinfo {author} {\bibfnamefont {S.~J.}\ \bibnamefont {Bradshaw}}, \ and\ \bibinfo {author} {\bibfnamefont {T.~C.}\ \bibnamefont {Killian}},\ }\href {\doibase 10.1103/PhysRevA.105.013108} {\bibfield  {journal} {\bibinfo  {journal} {Phys. Rev. A}\ }\textbf {\bibinfo {volume} {105}},\ \bibinfo {pages} {013108} (\bibinfo {year} {2022})}\BibitemShut {NoStop}%
\bibitem [{\citenamefont {Gorman}\ \emph {et~al.}(2021)\citenamefont {Gorman}, \citenamefont {Warrens}, \citenamefont {Bradshaw},\ and\ \citenamefont {Killian}}]{Gorman_PRL_2021}%
  \BibitemOpen
  \bibfield  {author} {\bibinfo {author} {\bibfnamefont {G.~M.}\ \bibnamefont {Gorman}}, \bibinfo {author} {\bibfnamefont {M.~K.}\ \bibnamefont {Warrens}}, \bibinfo {author} {\bibfnamefont {S.~J.}\ \bibnamefont {Bradshaw}}, \ and\ \bibinfo {author} {\bibfnamefont {T.~C.}\ \bibnamefont {Killian}},\ }\href {\doibase 10.1103/PhysRevLett.126.085002} {\bibfield  {journal} {\bibinfo  {journal} {Phys. Rev. Lett.}\ }\textbf {\bibinfo {volume} {126}},\ \bibinfo {pages} {085002} (\bibinfo {year} {2021})}\BibitemShut {NoStop}%
\bibitem [{\citenamefont {Zhang}\ \emph {et~al.}(2008)\citenamefont {Zhang}, \citenamefont {Fletcher}, \citenamefont {Rolston}, \citenamefont {Guzdar},\ and\ \citenamefont {Swisdak}}]{Zhang_PRL_2008}%
  \BibitemOpen
  \bibfield  {author} {\bibinfo {author} {\bibfnamefont {X.~L.}\ \bibnamefont {Zhang}}, \bibinfo {author} {\bibfnamefont {R.~S.}\ \bibnamefont {Fletcher}}, \bibinfo {author} {\bibfnamefont {S.~L.}\ \bibnamefont {Rolston}}, \bibinfo {author} {\bibfnamefont {P.~N.}\ \bibnamefont {Guzdar}}, \ and\ \bibinfo {author} {\bibfnamefont {M.}~\bibnamefont {Swisdak}},\ }\href {\doibase 10.1103/PhysRevLett.100.235002} {\bibfield  {journal} {\bibinfo  {journal} {Phys. Rev. Lett.}\ }\textbf {\bibinfo {volume} {100}},\ \bibinfo {pages} {235002} (\bibinfo {year} {2008})}\BibitemShut {NoStop}%
\bibitem [{\citenamefont {Sprenkle}\ \emph {et~al.}(2022)\citenamefont {Sprenkle}, \citenamefont {Bergeson}, \citenamefont {Silvestri},\ and\ \citenamefont {Murillo}}]{Sprenkle_PRE_2022}%
  \BibitemOpen
  \bibfield  {author} {\bibinfo {author} {\bibfnamefont {R.~T.}\ \bibnamefont {Sprenkle}}, \bibinfo {author} {\bibfnamefont {S.~D.}\ \bibnamefont {Bergeson}}, \bibinfo {author} {\bibfnamefont {L.~G.}\ \bibnamefont {Silvestri}}, \ and\ \bibinfo {author} {\bibfnamefont {M.~S.}\ \bibnamefont {Murillo}},\ }\href {\doibase 10.1103/PhysRevE.105.045201} {\bibfield  {journal} {\bibinfo  {journal} {Phys. Rev. E}\ }\textbf {\bibinfo {volume} {105}},\ \bibinfo {pages} {045201} (\bibinfo {year} {2022})}\BibitemShut {NoStop}%
\bibitem [{\citenamefont {Guthrie}\ and\ \citenamefont {Roberts}(2021)}]{Guthrie_POP_2021}%
  \BibitemOpen
  \bibfield  {author} {\bibinfo {author} {\bibfnamefont {J.~M.}\ \bibnamefont {Guthrie}}\ and\ \bibinfo {author} {\bibfnamefont {J.~L.}\ \bibnamefont {Roberts}},\ }\href {\doibase 10.1063/5.0047640} {\bibfield  {journal} {\bibinfo  {journal} {Physics of Plasmas}\ }\textbf {\bibinfo {volume} {28}},\ \bibinfo {pages} {052101} (\bibinfo {year} {2021})},\ \Eprint {http://arxiv.org/abs/https://doi.org/10.1063/5.0047640} {https://doi.org/10.1063/5.0047640} \BibitemShut {NoStop}%
\bibitem [{\citenamefont {Pak}\ \emph {et~al.}(2024)\citenamefont {Pak}, \citenamefont {Billings}, \citenamefont {Schlitters}, \citenamefont {Bergeson},\ and\ \citenamefont {Murillo}}]{Pak_PRE_2024}%
  \BibitemOpen
  \bibfield  {author} {\bibinfo {author} {\bibfnamefont {C.}~\bibnamefont {Pak}}, \bibinfo {author} {\bibfnamefont {V.}~\bibnamefont {Billings}}, \bibinfo {author} {\bibfnamefont {M.}~\bibnamefont {Schlitters}}, \bibinfo {author} {\bibfnamefont {S.~D.}\ \bibnamefont {Bergeson}}, \ and\ \bibinfo {author} {\bibfnamefont {M.~S.}\ \bibnamefont {Murillo}},\ }\href {\doibase 10.1103/PhysRevE.109.015201} {\bibfield  {journal} {\bibinfo  {journal} {Phys. Rev. E}\ }\textbf {\bibinfo {volume} {109}},\ \bibinfo {pages} {015201} (\bibinfo {year} {2024})}\BibitemShut {NoStop}%
\bibitem [{\citenamefont {Bennett}\ \emph {et~al.}(2021)\citenamefont {Bennett}, \citenamefont {Welch}, \citenamefont {Laity}, \citenamefont {Rose},\ and\ \citenamefont {Cuneo}}]{Bennett_Phys_Rev_Accel_Beams_2021}%
  \BibitemOpen
  \bibfield  {author} {\bibinfo {author} {\bibfnamefont {N.}~\bibnamefont {Bennett}}, \bibinfo {author} {\bibfnamefont {D.~R.}\ \bibnamefont {Welch}}, \bibinfo {author} {\bibfnamefont {G.}~\bibnamefont {Laity}}, \bibinfo {author} {\bibfnamefont {D.~V.}\ \bibnamefont {Rose}}, \ and\ \bibinfo {author} {\bibfnamefont {M.~E.}\ \bibnamefont {Cuneo}},\ }\href {\doibase 10.1103/PhysRevAccelBeams.24.060401} {\bibfield  {journal} {\bibinfo  {journal} {Phys. Rev. Accel. Beams}\ }\textbf {\bibinfo {volume} {24}},\ \bibinfo {pages} {060401} (\bibinfo {year} {2021})}\BibitemShut {NoStop}%
\bibitem [{\citenamefont {Jose}\ and\ \citenamefont {Baalrud}(2020)}]{Jose_POP_2020}%
  \BibitemOpen
  \bibfield  {author} {\bibinfo {author} {\bibfnamefont {L.}~\bibnamefont {Jose}}\ and\ \bibinfo {author} {\bibfnamefont {S.~D.}\ \bibnamefont {Baalrud}},\ }\href {\doibase 10.1063/5.0025158} {\bibfield  {journal} {\bibinfo  {journal} {Physics of Plasmas}\ }\textbf {\bibinfo {volume} {27}},\ \bibinfo {pages} {112101} (\bibinfo {year} {2020})}\BibitemShut {NoStop}%
\bibitem [{\citenamefont {Jose}\ \emph {et~al.}(2022)\citenamefont {Jose}, \citenamefont {Bernstein},\ and\ \citenamefont {Baalrud}}]{Jose_POP_2022}%
  \BibitemOpen
  \bibfield  {author} {\bibinfo {author} {\bibfnamefont {L.}~\bibnamefont {Jose}}, \bibinfo {author} {\bibfnamefont {D.~J.}\ \bibnamefont {Bernstein}}, \ and\ \bibinfo {author} {\bibfnamefont {S.~D.}\ \bibnamefont {Baalrud}},\ }\href {\doibase 10.1063/5.0121285} {\bibfield  {journal} {\bibinfo  {journal} {Physics of Plasmas}\ }\textbf {\bibinfo {volume} {29}},\ \bibinfo {pages} {112103} (\bibinfo {year} {2022})},\ \Eprint {http://arxiv.org/abs/https://doi.org/10.1063/5.0121285} {https://doi.org/10.1063/5.0121285} \BibitemShut {NoStop}%
\bibitem [{\citenamefont {Jose}\ and\ \citenamefont {Baalrud}(2023)}]{Jose_POP_2023}%
  \BibitemOpen
  \bibfield  {author} {\bibinfo {author} {\bibfnamefont {L.}~\bibnamefont {Jose}}\ and\ \bibinfo {author} {\bibfnamefont {S.~D.}\ \bibnamefont {Baalrud}},\ }\href {\doibase 10.1063/5.0146417} {\bibfield  {journal} {\bibinfo  {journal} {Physics of Plasmas}\ }\textbf {\bibinfo {volume} {30}},\ \bibinfo {pages} {052103} (\bibinfo {year} {2023})}\BibitemShut {NoStop}%
\bibitem [{\citenamefont {Ferziger}\ and\ \citenamefont {Kaper}(1972)}]{ferziger1972mathematical}%
  \BibitemOpen
  \bibfield  {author} {\bibinfo {author} {\bibfnamefont {J.~H.}\ \bibnamefont {Ferziger}}\ and\ \bibinfo {author} {\bibfnamefont {H.~G.}\ \bibnamefont {Kaper}},\ }\href@noop {} {\emph {\bibinfo {title} {Mathematical theory of transport processes in gases}}}\ (\bibinfo  {publisher} {North-Holland},\ \bibinfo {year} {1972})\BibitemShut {NoStop}%
\bibitem [{\citenamefont {Rosenbluth}\ \emph {et~al.}(1957)\citenamefont {Rosenbluth}, \citenamefont {MacDonald},\ and\ \citenamefont {Judd}}]{Rosenbluth_PR_1957}%
  \BibitemOpen
  \bibfield  {author} {\bibinfo {author} {\bibfnamefont {M.~N.}\ \bibnamefont {Rosenbluth}}, \bibinfo {author} {\bibfnamefont {W.~M.}\ \bibnamefont {MacDonald}}, \ and\ \bibinfo {author} {\bibfnamefont {D.~L.}\ \bibnamefont {Judd}},\ }\href {\doibase 10.1103/PhysRev.107.1} {\bibfield  {journal} {\bibinfo  {journal} {Phys. Rev.}\ }\textbf {\bibinfo {volume} {107}},\ \bibinfo {pages} {1} (\bibinfo {year} {1957})}\BibitemShut {NoStop}%
\bibitem [{\citenamefont {Lenard}(1960)}]{Lenard_AP_1960}%
  \BibitemOpen
  \bibfield  {author} {\bibinfo {author} {\bibfnamefont {A.}~\bibnamefont {Lenard}},\ }\href {\doibase https://doi.org/10.1016/0003-4916(60)90003-8} {\bibfield  {journal} {\bibinfo  {journal} {Annals of Physics}\ }\textbf {\bibinfo {volume} {10}},\ \bibinfo {pages} {390} (\bibinfo {year} {1960})}\BibitemShut {NoStop}%
\bibitem [{\citenamefont {Balescu}(1960)}]{Balescu_Phys_Fluids_1960}%
  \BibitemOpen
  \bibfield  {author} {\bibinfo {author} {\bibfnamefont {R.}~\bibnamefont {Balescu}},\ }\href {\doibase 10.1063/1.1706002} {\bibfield  {journal} {\bibinfo  {journal} {The Physics of Fluids}\ }\textbf {\bibinfo {volume} {3}},\ \bibinfo {pages} {52} (\bibinfo {year} {1960})},\ \Eprint {http://arxiv.org/abs/https://aip.scitation.org/doi/pdf/10.1063/1.1706002} {https://aip.scitation.org/doi/pdf/10.1063/1.1706002} \BibitemShut {NoStop}%
\bibitem [{\citenamefont {Jose}\ and\ \citenamefont {Baalrud}(2021)}]{Jose_POP_2021}%
  \BibitemOpen
  \bibfield  {author} {\bibinfo {author} {\bibfnamefont {L.}~\bibnamefont {Jose}}\ and\ \bibinfo {author} {\bibfnamefont {S.~D.}\ \bibnamefont {Baalrud}},\ }\href {\doibase 10.1063/5.0054552} {\bibfield  {journal} {\bibinfo  {journal} {Physics of Plasmas}\ }\textbf {\bibinfo {volume} {28}},\ \bibinfo {pages} {072107} (\bibinfo {year} {2021})},\ \Eprint {http://arxiv.org/abs/https://doi.org/10.1063/5.0054552} {https://doi.org/10.1063/5.0054552} \BibitemShut {NoStop}%
\bibitem [{\citenamefont {Baalrud}\ and\ \citenamefont {Daligault}(2013)}]{Baalrud_PRL_2013}%
  \BibitemOpen
  \bibfield  {author} {\bibinfo {author} {\bibfnamefont {S.~D.}\ \bibnamefont {Baalrud}}\ and\ \bibinfo {author} {\bibfnamefont {J.}~\bibnamefont {Daligault}},\ }\href {\doibase 10.1103/PhysRevLett.110.235001} {\bibfield  {journal} {\bibinfo  {journal} {Phys. Rev. Lett.}\ }\textbf {\bibinfo {volume} {110}},\ \bibinfo {pages} {235001} (\bibinfo {year} {2013})}\BibitemShut {NoStop}%
\bibitem [{\citenamefont {Baalrud}\ and\ \citenamefont {Daligault}(2019)}]{Baalrud_POP_2019}%
  \BibitemOpen
  \bibfield  {author} {\bibinfo {author} {\bibfnamefont {S.~D.}\ \bibnamefont {Baalrud}}\ and\ \bibinfo {author} {\bibfnamefont {J.}~\bibnamefont {Daligault}},\ }\href@noop {} {\bibfield  {journal} {\bibinfo  {journal} {Physics of Plasmas}\ }\textbf {\bibinfo {volume} {26}},\ \bibinfo {pages} {082106} (\bibinfo {year} {2019})}\BibitemShut {NoStop}%
\bibitem [{\citenamefont {Baalrud}\ and\ \citenamefont {Daligault}(2014)}]{Baalrud_POP_2014}%
  \BibitemOpen
  \bibfield  {author} {\bibinfo {author} {\bibfnamefont {S.~D.}\ \bibnamefont {Baalrud}}\ and\ \bibinfo {author} {\bibfnamefont {J.}~\bibnamefont {Daligault}},\ }\href {\doibase 10.1063/1.4875282} {\bibfield  {journal} {\bibinfo  {journal} {Physics of Plasmas}\ }\textbf {\bibinfo {volume} {21}},\ \bibinfo {pages} {055707} (\bibinfo {year} {2014})}\BibitemShut {NoStop}%
\bibitem [{\citenamefont {Lafleur}\ and\ \citenamefont {Baalrud}(2019)}]{Lafleur_PPCF_2019}%
  \BibitemOpen
  \bibfield  {author} {\bibinfo {author} {\bibfnamefont {T.}~\bibnamefont {Lafleur}}\ and\ \bibinfo {author} {\bibfnamefont {S.~D.}\ \bibnamefont {Baalrud}},\ }\href@noop {} {\bibfield  {journal} {\bibinfo  {journal} {Plasma Physics and Controlled Fusion}\ }\textbf {\bibinfo {volume} {61}},\ \bibinfo {pages} {125004} (\bibinfo {year} {2019})}\BibitemShut {NoStop}%
\bibitem [{\citenamefont {Lafleur}\ and\ \citenamefont {Baalrud}(2020)}]{Lafleur_PPCF_2020}%
  \BibitemOpen
  \bibfield  {author} {\bibinfo {author} {\bibfnamefont {T.}~\bibnamefont {Lafleur}}\ and\ \bibinfo {author} {\bibfnamefont {S.~D.}\ \bibnamefont {Baalrud}},\ }\href {\doibase 10.1088/1361-6587/ab9bea} {\bibfield  {journal} {\bibinfo  {journal} {Plasma Physics and Controlled Fusion}\ }\textbf {\bibinfo {volume} {62}},\ \bibinfo {pages} {095003} (\bibinfo {year} {2020})}\BibitemShut {NoStop}%
\bibitem [{\citenamefont {Bernstein}\ \emph {et~al.}(2020)\citenamefont {Bernstein}, \citenamefont {Lafleur}, \citenamefont {Daligault},\ and\ \citenamefont {Baalrud}}]{David_PRE_2020}%
  \BibitemOpen
  \bibfield  {author} {\bibinfo {author} {\bibfnamefont {D.~J.}\ \bibnamefont {Bernstein}}, \bibinfo {author} {\bibfnamefont {T.}~\bibnamefont {Lafleur}}, \bibinfo {author} {\bibfnamefont {J.}~\bibnamefont {Daligault}}, \ and\ \bibinfo {author} {\bibfnamefont {S.~D.}\ \bibnamefont {Baalrud}},\ }\href {\doibase 10.1103/PhysRevE.102.041201} {\bibfield  {journal} {\bibinfo  {journal} {Phys. Rev. E}\ }\textbf {\bibinfo {volume} {102}},\ \bibinfo {pages} {041201} (\bibinfo {year} {2020})}\BibitemShut {NoStop}%
\bibitem [{\citenamefont {Baalrud}\ and\ \citenamefont {Lafleur}(2021)}]{Baalrud_POP_2021}%
  \BibitemOpen
  \bibfield  {author} {\bibinfo {author} {\bibfnamefont {S.~D.}\ \bibnamefont {Baalrud}}\ and\ \bibinfo {author} {\bibfnamefont {T.}~\bibnamefont {Lafleur}},\ }\href {\doibase 10.1063/5.0054113} {\bibfield  {journal} {\bibinfo  {journal} {Physics of Plasmas}\ }\textbf {\bibinfo {volume} {28}},\ \bibinfo {pages} {102107} (\bibinfo {year} {2021})},\ \Eprint {http://arxiv.org/abs/https://doi.org/10.1063/5.0054113} {https://doi.org/10.1063/5.0054113} \BibitemShut {NoStop}%
\bibitem [{\citenamefont {Silin}(1963)}]{Silin_1963}%
  \BibitemOpen
  \bibfield  {author} {\bibinfo {author} {\bibfnamefont {V.}~\bibnamefont {Silin}},\ }\href@noop {} {\bibfield  {journal} {\bibinfo  {journal} {Sov. Phys. JETP}\ }\textbf {\bibinfo {volume} {16}},\ \bibinfo {pages} {1281} (\bibinfo {year} {1963})}\BibitemShut {NoStop}%
\bibitem [{\citenamefont {Kihara}\ \emph {et~al.}(1960)\citenamefont {Kihara}, \citenamefont {Midzuno}, \citenamefont {Sakuma},\ and\ \citenamefont {Shizume}}]{Kihara_JPSJ_1960}%
  \BibitemOpen
  \bibfield  {author} {\bibinfo {author} {\bibfnamefont {T.}~\bibnamefont {Kihara}}, \bibinfo {author} {\bibfnamefont {Y.}~\bibnamefont {Midzuno}}, \bibinfo {author} {\bibfnamefont {K.}~\bibnamefont {Sakuma}}, \ and\ \bibinfo {author} {\bibfnamefont {T.}~\bibnamefont {Shizume}},\ }\href {\doibase 10.1143/JPSJ.15.684} {\bibfield  {journal} {\bibinfo  {journal} {Journal of the Physical Society of Japan}\ }\textbf {\bibinfo {volume} {15}},\ \bibinfo {pages} {684} (\bibinfo {year} {1960})},\ \Eprint {http://arxiv.org/abs/https://doi.org/10.1143/JPSJ.15.684} {https://doi.org/10.1143/JPSJ.15.684} \BibitemShut {NoStop}%
\bibitem [{\citenamefont {Kihara}\ and\ \citenamefont {Midzuno}(1960)}]{Kihara_RevModPhys_1960}%
  \BibitemOpen
  \bibfield  {author} {\bibinfo {author} {\bibfnamefont {T.}~\bibnamefont {Kihara}}\ and\ \bibinfo {author} {\bibfnamefont {Y.}~\bibnamefont {Midzuno}},\ }\href {\doibase 10.1103/RevModPhys.32.722} {\bibfield  {journal} {\bibinfo  {journal} {Rev. Mod. Phys.}\ }\textbf {\bibinfo {volume} {32}},\ \bibinfo {pages} {722} (\bibinfo {year} {1960})}\BibitemShut {NoStop}%
\bibitem [{\citenamefont {Ichimaru}\ and\ \citenamefont {Rosenbluth}(1970)}]{Ichimaru_Phy_Fluids_1970}%
  \BibitemOpen
  \bibfield  {author} {\bibinfo {author} {\bibfnamefont {S.}~\bibnamefont {Ichimaru}}\ and\ \bibinfo {author} {\bibfnamefont {M.~N.}\ \bibnamefont {Rosenbluth}},\ }\href {\doibase 10.1063/1.1692864} {\bibfield  {journal} {\bibinfo  {journal} {The Physics of Fluids}\ }\textbf {\bibinfo {volume} {13}},\ \bibinfo {pages} {2778} (\bibinfo {year} {1970})},\ \Eprint {http://arxiv.org/abs/https://aip.scitation.org/doi/pdf/10.1063/1.1692864} {https://aip.scitation.org/doi/pdf/10.1063/1.1692864} \BibitemShut {NoStop}%
\bibitem [{\citenamefont {Nersisyan}\ \emph {et~al.}(2011)\citenamefont {Nersisyan}, \citenamefont {Deutsch},\ and\ \citenamefont {Das}}]{Nersisyan_PRE_2011}%
  \BibitemOpen
  \bibfield  {author} {\bibinfo {author} {\bibfnamefont {H.~B.}\ \bibnamefont {Nersisyan}}, \bibinfo {author} {\bibfnamefont {C.}~\bibnamefont {Deutsch}}, \ and\ \bibinfo {author} {\bibfnamefont {A.~K.}\ \bibnamefont {Das}},\ }\href {\doibase 10.1103/PhysRevE.83.036403} {\bibfield  {journal} {\bibinfo  {journal} {Phys. Rev. E}\ }\textbf {\bibinfo {volume} {83}},\ \bibinfo {pages} {036403} (\bibinfo {year} {2011})}\BibitemShut {NoStop}%
\bibitem [{\citenamefont {Dong}\ \emph {et~al.}(2013{\natexlab{a}})\citenamefont {Dong}, \citenamefont {Ren}, \citenamefont {Cai},\ and\ \citenamefont {Li}}]{Dong_POP_2013}%
  \BibitemOpen
  \bibfield  {author} {\bibinfo {author} {\bibfnamefont {C.}~\bibnamefont {Dong}}, \bibinfo {author} {\bibfnamefont {H.}~\bibnamefont {Ren}}, \bibinfo {author} {\bibfnamefont {H.}~\bibnamefont {Cai}}, \ and\ \bibinfo {author} {\bibfnamefont {D.}~\bibnamefont {Li}},\ }\href@noop {} {\bibfield  {journal} {\bibinfo  {journal} {Physics of Plasmas}\ }\textbf {\bibinfo {volume} {20}},\ \bibinfo {pages} {032512} (\bibinfo {year} {2013}{\natexlab{a}})}\BibitemShut {NoStop}%
\bibitem [{\citenamefont {Dong}\ \emph {et~al.}(2013{\natexlab{b}})\citenamefont {Dong}, \citenamefont {Ren}, \citenamefont {Cai},\ and\ \citenamefont {Li}}]{Dong_POP_2013_2}%
  \BibitemOpen
  \bibfield  {author} {\bibinfo {author} {\bibfnamefont {C.}~\bibnamefont {Dong}}, \bibinfo {author} {\bibfnamefont {H.}~\bibnamefont {Ren}}, \bibinfo {author} {\bibfnamefont {H.}~\bibnamefont {Cai}}, \ and\ \bibinfo {author} {\bibfnamefont {D.}~\bibnamefont {Li}},\ }\href@noop {} {\bibfield  {journal} {\bibinfo  {journal} {Physics of Plasmas}\ }\textbf {\bibinfo {volume} {20}},\ \bibinfo {pages} {102518} (\bibinfo {year} {2013}{\natexlab{b}})}\BibitemShut {NoStop}%
\bibitem [{\citenamefont {Kihara}(1959)}]{Kihara_JPSJ_1959}%
  \BibitemOpen
  \bibfield  {author} {\bibinfo {author} {\bibfnamefont {T.}~\bibnamefont {Kihara}},\ }\href {\doibase 10.1143/JPSJ.14.1751} {\bibfield  {journal} {\bibinfo  {journal} {Journal of the Physical Society of Japan}\ }\textbf {\bibinfo {volume} {14}},\ \bibinfo {pages} {1751} (\bibinfo {year} {1959})},\ \Eprint {http://arxiv.org/abs/https://doi.org/10.1143/JPSJ.14.1751} {https://doi.org/10.1143/JPSJ.14.1751} \BibitemShut {NoStop}%
\bibitem [{\citenamefont {Baalrud}(2012)}]{baalrud2012transport}%
  \BibitemOpen
  \bibfield  {author} {\bibinfo {author} {\bibfnamefont {S.~D.}\ \bibnamefont {Baalrud}},\ }\href@noop {} {\bibfield  {journal} {\bibinfo  {journal} {Physics of Plasmas}\ }\textbf {\bibinfo {volume} {19}},\ \bibinfo {pages} {030701} (\bibinfo {year} {2012})}\BibitemShut {NoStop}%
\bibitem [{\citenamefont {Ichimaru}(2004)}]{ichimaru2018statistical}%
  \BibitemOpen
  \bibfield  {author} {\bibinfo {author} {\bibfnamefont {S.}~\bibnamefont {Ichimaru}},\ }\href@noop {} {\emph {\bibinfo {title} {Statistical Plasma Physics, Volume I: Basic Principles}}}\ (\bibinfo  {publisher} {CRC Press},\ \bibinfo {year} {2004})\BibitemShut {NoStop}%
\bibitem [{\citenamefont {Nicholson}(1983)}]{nicholson1983introduction}%
  \BibitemOpen
  \bibfield  {author} {\bibinfo {author} {\bibfnamefont {D.~R.}\ \bibnamefont {Nicholson}},\ }\href@noop {} {\emph {\bibinfo {title} {Introduction to plasma theory}}}\ (\bibinfo  {publisher} {Wiley},\ \bibinfo {year} {1983})\BibitemShut {NoStop}%
\bibitem [{\citenamefont {Hansen}\ and\ \citenamefont {McDonald}(2013)}]{hansen2013theory}%
  \BibitemOpen
  \bibfield  {author} {\bibinfo {author} {\bibfnamefont {J.~P.}\ \bibnamefont {Hansen}}\ and\ \bibinfo {author} {\bibfnamefont {I.~R.}\ \bibnamefont {McDonald}},\ }\href@noop {} {\emph {\bibinfo {title} {Theory of simple liquids: with applications to soft matter}}}\ (\bibinfo  {publisher} {Academic Press},\ \bibinfo {year} {2013})\BibitemShut {NoStop}%
\bibitem [{\citenamefont {Jose}(2023)}]{jose2023kinetic}%
  \BibitemOpen
  \bibfield  {author} {\bibinfo {author} {\bibfnamefont {L.}~\bibnamefont {Jose}},\ }\emph {\bibinfo {title} {Kinetic Theory of Strongly Magnetized Plasmas}},\ \href@noop {} {Ph.D. thesis},\ \bibinfo  {school} {University of Michigan} (\bibinfo {year} {2023})\BibitemShut {NoStop}%
\bibitem [{\citenamefont {Hairer}\ \emph {et~al.}(1993)\citenamefont {Hairer}, \citenamefont {Norsett},\ and\ \citenamefont {Wanner}}]{DOP853}%
  \BibitemOpen
  \bibfield  {author} {\bibinfo {author} {\bibfnamefont {E.}~\bibnamefont {Hairer}}, \bibinfo {author} {\bibfnamefont {S.~P.}\ \bibnamefont {Norsett}}, \ and\ \bibinfo {author} {\bibfnamefont {G.}~\bibnamefont {Wanner}},\ }\enquote {\bibinfo {title} {Runge-kutta and extrapolation methods},}\ in\ \href {\doibase 10.1007/978-3-540-78862-1_2} {\emph {\bibinfo {booktitle} {Solving Ordinary Differential Equations I: Nonstiff Problems}}}\ (\bibinfo  {publisher} {Springer Berlin Heidelberg},\ \bibinfo {year} {1993})\ pp.\ \bibinfo {pages} {129--353}\BibitemShut {NoStop}%
\bibitem [{\citenamefont {Lepage}(2021)}]{LEPAGE_2021_JCP}%
  \BibitemOpen
  \bibfield  {author} {\bibinfo {author} {\bibfnamefont {G.~P.}\ \bibnamefont {Lepage}},\ }\href {\doibase https://doi.org/10.1016/j.jcp.2021.110386} {\bibfield  {journal} {\bibinfo  {journal} {Journal of Computational Physics}\ }\textbf {\bibinfo {volume} {439}},\ \bibinfo {pages} {110386} (\bibinfo {year} {2021})}\BibitemShut {NoStop}%
\bibitem [{\citenamefont {Lepage}(2024)}]{peter_lepage_2024_12687656}%
  \BibitemOpen
  \bibfield  {author} {\bibinfo {author} {\bibfnamefont {P.}~\bibnamefont {Lepage}},\ }\href {\doibase 10.5281/zenodo.12687656} {\enquote {\bibinfo {title} {gplepage/vegas: vegas version 6.1.3},}\ } (\bibinfo {year} {2024})\BibitemShut {NoStop}%
\bibitem [{\citenamefont {O’Neil}(1983)}]{Oneil_Phys_Fluids_1983}%
  \BibitemOpen
  \bibfield  {author} {\bibinfo {author} {\bibfnamefont {T.}~\bibnamefont {O’Neil}},\ }\href@noop {} {\bibfield  {journal} {\bibinfo  {journal} {Physics of Fluids}\ }\textbf {\bibinfo {volume} {26}},\ \bibinfo {pages} {2128} (\bibinfo {year} {1983})}\BibitemShut {NoStop}%
\bibitem [{\citenamefont {O’Neil}\ and\ \citenamefont {Hjorth}(1985)}]{Oneil_Phys_Fluids_1985}%
  \BibitemOpen
  \bibfield  {author} {\bibinfo {author} {\bibfnamefont {T.}~\bibnamefont {O’Neil}}\ and\ \bibinfo {author} {\bibfnamefont {P.}~\bibnamefont {Hjorth}},\ }\href@noop {} {\bibfield  {journal} {\bibinfo  {journal} {Physics of fluids}\ }\textbf {\bibinfo {volume} {28}},\ \bibinfo {pages} {3241} (\bibinfo {year} {1985})}\BibitemShut {NoStop}%
\bibitem [{\citenamefont {Andresen}\ \emph {et~al.}(2008)\citenamefont {Andresen}, \citenamefont {Bertsche}, \citenamefont {Bowe}, \citenamefont {Bray}, \citenamefont {Butler}, \citenamefont {Cesar}, \citenamefont {Chapman}, \citenamefont {Charlton}, \citenamefont {Fajans}, \citenamefont {Fujiwara}, \citenamefont {Funakoshi}, \citenamefont {Gill}, \citenamefont {Hangst}, \citenamefont {Hardy}, \citenamefont {Hayano}, \citenamefont {Hayden}, \citenamefont {Hydomako}, \citenamefont {Jenkins}, \citenamefont {Jorgensen}, \citenamefont {Kurchaninov}, \citenamefont {Lambo}, \citenamefont {Madsen}, \citenamefont {Nolan}, \citenamefont {Olchanski}, \citenamefont {Olin}, \citenamefont {Povilus}, \citenamefont {Pusa}, \citenamefont {Robicheaux}, \citenamefont {Sarid}, \citenamefont {Seif El~Nasr}, \citenamefont {Silveira}, \citenamefont {Storey}, \citenamefont {Thompson}, \citenamefont {van~der Werf}, \citenamefont {Wurtele},\ and\ \citenamefont {Yamazaki}}]{Andresen_PRL_2008}%
  \BibitemOpen
  \bibfield  {author} {\bibinfo {author} {\bibfnamefont {G.~B.}\ \bibnamefont {Andresen}}, \bibinfo {author} {\bibfnamefont {W.}~\bibnamefont {Bertsche}}, \bibinfo {author} {\bibfnamefont {P.~D.}\ \bibnamefont {Bowe}}, \bibinfo {author} {\bibfnamefont {C.~C.}\ \bibnamefont {Bray}}, \bibinfo {author} {\bibfnamefont {E.}~\bibnamefont {Butler}}, \bibinfo {author} {\bibfnamefont {C.~L.}\ \bibnamefont {Cesar}}, \bibinfo {author} {\bibfnamefont {S.}~\bibnamefont {Chapman}}, \bibinfo {author} {\bibfnamefont {M.}~\bibnamefont {Charlton}}, \bibinfo {author} {\bibfnamefont {J.}~\bibnamefont {Fajans}}, \bibinfo {author} {\bibfnamefont {M.~C.}\ \bibnamefont {Fujiwara}}, \bibinfo {author} {\bibfnamefont {R.}~\bibnamefont {Funakoshi}}, \bibinfo {author} {\bibfnamefont {D.~R.}\ \bibnamefont {Gill}}, \bibinfo {author} {\bibfnamefont {J.~S.}\ \bibnamefont {Hangst}}, \bibinfo {author} {\bibfnamefont {W.~N.}\ \bibnamefont {Hardy}}, \bibinfo {author} {\bibfnamefont {R.~S.}\ \bibnamefont {Hayano}}, \bibinfo {author}
  {\bibfnamefont {M.~E.}\ \bibnamefont {Hayden}}, \bibinfo {author} {\bibfnamefont {R.}~\bibnamefont {Hydomako}}, \bibinfo {author} {\bibfnamefont {M.~J.}\ \bibnamefont {Jenkins}}, \bibinfo {author} {\bibfnamefont {L.~V.}\ \bibnamefont {Jorgensen}}, \bibinfo {author} {\bibfnamefont {L.}~\bibnamefont {Kurchaninov}}, \bibinfo {author} {\bibfnamefont {R.}~\bibnamefont {Lambo}}, \bibinfo {author} {\bibfnamefont {N.}~\bibnamefont {Madsen}}, \bibinfo {author} {\bibfnamefont {P.}~\bibnamefont {Nolan}}, \bibinfo {author} {\bibfnamefont {K.}~\bibnamefont {Olchanski}}, \bibinfo {author} {\bibfnamefont {A.}~\bibnamefont {Olin}}, \bibinfo {author} {\bibfnamefont {A.}~\bibnamefont {Povilus}}, \bibinfo {author} {\bibfnamefont {P.}~\bibnamefont {Pusa}}, \bibinfo {author} {\bibfnamefont {F.}~\bibnamefont {Robicheaux}}, \bibinfo {author} {\bibfnamefont {E.}~\bibnamefont {Sarid}}, \bibinfo {author} {\bibfnamefont {S.}~\bibnamefont {Seif El~Nasr}}, \bibinfo {author} {\bibfnamefont {D.~M.}\ \bibnamefont {Silveira}}, \bibinfo
  {author} {\bibfnamefont {J.~W.}\ \bibnamefont {Storey}}, \bibinfo {author} {\bibfnamefont {R.~I.}\ \bibnamefont {Thompson}}, \bibinfo {author} {\bibfnamefont {D.~P.}\ \bibnamefont {van~der Werf}}, \bibinfo {author} {\bibfnamefont {J.~S.}\ \bibnamefont {Wurtele}}, \ and\ \bibinfo {author} {\bibfnamefont {Y.}~\bibnamefont {Yamazaki}},\ }\href {\doibase 10.1103/PhysRevLett.100.203401} {\bibfield  {journal} {\bibinfo  {journal} {Phys. Rev. Lett.}\ }\textbf {\bibinfo {volume} {100}},\ \bibinfo {pages} {203401} (\bibinfo {year} {2008})}\BibitemShut {NoStop}%
\end{thebibliography}%

\end{document}